\documentclass[preprint,showpacs,preprintnumbers,amsmath,amssymb]{revtex4}
\usepackage{latexsym,amsmath,tabularx,amssymb,bm}
\usepackage{epsfig}
\usepackage{citesort}

\long\def\comment#1{ }

\usepackage{amssymb}
\usepackage{makeidx}
\usepackage{graphicx}

\def\simge{\mathrel{%
    \rlap{\raise 0.511ex \hbox{$>$}}{\lower 0.511ex \hbox{$\sim$}}}}
\def\simle{\mathrel{
    \rlap{\raise 0.511ex \hbox{$<$}}{\lower 0.511ex \hbox{$\sim$}}}}
\newcommand{\beq}{\begin{eqnarray}}
\newcommand{\eeq}{\end{eqnarray}}
\newcommand{\del}{\partial}

\begin{document}
\preprint{ ECT*-- 05-26}
\title{Non-Perturbative Renormalization Group calculation
of the scalar self-energy}
\author{Jean-Paul Blaizot}
  \thanks{Member of CNRS} \email{blaizot@ect.it}
  \affiliation{ECT*, Villa Tambosi, strada delle Tabarelle 286, 38050 Villazzano (TN), Italy}
  \author{Ram\'on M\'endez-Galain}
  \email{mendezg@fing.edu.uy}
  \affiliation{Instituto de F\'{\i}sica, Facultad de Ingenier\'{\i}a, J.H.y Reissig 565, 11000
  Montevideo, Uruguay}
\author{Nicol\'as Wschebor}
  \email{nicws@fing.edu.uy}
  \affiliation{
  Instituto de F\'{\i}sica, Facultad de Ingenier\'{\i}a, J.H.y Reissig 565, 11000
  Montevideo, Uruguay}

  \date{\today}
\begin{abstract}
We present the first numerical application of a method that we
have recently proposed to solve the Non Perturbative Renormalization
Group equations and obtain the $n$-point functions for arbitrary
external momenta. This method leads to flow equations for the
$n$-point functions  which are also differential equations with
respect to a constant background field. This makes them,  a priori,
difficult to solve. However, we demonstrate in this paper that,
within  a simple approximation which turns out to be quite accurate,
the solution of these flow equations is not more complicated than that
 of the flow equations  obtained in the
derivative expansion. Thus, with a numerical effort comparable to
that involved in the derivative expansion, we can get the full
momentum dependence of the $n$-point functions. The method is
applied, in its leading order, to the calculation of the self-energy
in a 3-dimensional scalar field theory, at criticality. Accurate
results are obtained over the entire range of momenta.
\end{abstract}

\maketitle

\section{Introduction}

The non perturbative renormalization group (NPRG)
\cite{Wetterich93,Ellwanger93,Tetradis94,Morris94,Morris94c}  stands
out as a very promising formalism to address non perturbative
problems, i.e., problems in which the absence of a small parameter
prevents one to build a solution in terms of a systematic expansion.
It leads to exact flow equations which are difficult to solve in
general, but which offer the possibility for new approximation
schemes. When only correlation functions at small momenta are
needed, as is the case for instance in the calculation of critical exponents, a general approximation method to solve the infinite
hierarchy of the NPRG equations has been developped
\cite{Morris94c,Bagnuls:2000ae,Berges02}. This method, which can be
systematically improved, is based on a derivative expansion of the
the effective action. It has been applied successfully to a variety
of physical problems, in condensed matter, particle or nuclear
physics (for  reviews, see e.g.
\cite{Bagnuls:2000ae,Berges02}). However, in many
situations, this is not enough:  in order to calculate the
quantities of physical interest, the knowledge of the full momentum
dependence of the correlation functions is mandatory. Many efforts
to get this information from  the flow equations, involve
truncations of various kinds \cite{truncation}, following an early
suggestion by Weinberg \cite{weinberg73} (see however \cite{Golner,Parola}).

The present paper explores the applicability of the strategy that we
proposed recently in \cite{Blaizot:2005xy}, following our previous
works \cite{Blaizot:2005wd,Blaizot:2006vr} in which we presented a
scheme to obtain the momentum dependence of $n$-point functions from
the flow equations. The strategy put forward in
\cite{Blaizot:2005xy} is based on the fact that the internal
momentum $q$ in the integrals that determine the flow of the
$n$-point functions is bounded by the regulator introduced by the
NPRG. Since this regulator also guarantees that the vertex functions
are smooth functions of the momenta, these can be expanded in powers
of $q^2/\kappa^2$, $\kappa$ being the cut-off scale in the
regulator. The  ``leading order'' (LO) of the approximation scheme
proposed in \cite{Blaizot:2005xy} simply consists in keeping the
lowest order of this expansion, i.e., in setting $q=0$ in the
vertices. Doing so, and working in a constant external field, it is
possible to relate to each other the various $n$-point functions
that appear in a given flow equation through derivatives with
respect to the external field, thereby closing the hierarchy of
equations.

In \cite{Blaizot:2005xy} we showed that the method  reproduces perturbative results, at any desired order. Furthermore, we also showed that the LO  is
exact in the large $N$ limit of the $O(N)$ scalar model. Finally, one
expects the method to provide, at each order of the expansion,
results as good as those of the derivative expansion in the domain where the derivative expansion is valid.

The price to pay is that the resulting equations are also
differential equations with respect to a uniform background field, with integral kernels that involve  the solution itself.
These integro-differential equations are a priori difficult to solve. 
The aim of this paper is to demonstrate that they can indeed be solved, with a numerical effort comparable to that involved in solving the flow equations that result from the derivative expansion, and to present a first  application to the 
study of the 2-point correlation function
of the
  scalar model, in the LO of the
approximation scheme. 

The outline of the paper is as follows. In section \ref{method}  we  briefly recall some basic features of the  NPRG  and the essence of our approximation scheme in the case of a scalar field theory. In
section \ref{analysis} we  analyze the structure of the 
flow equation for the 2-point correlation function and describe the
strategy that we used  to solve it. In section \ref{numerics} we present  numerical results for the self-energy of the scalar field, at criticality and in $d=3$.  The appendices gather technical material.

\section{The method}
\label{method} We  consider a scalar field theory with
the classical action \beq\label{eactON} S = \int {\rm
d}^{d}x\,\left\lbrace{ \frac{1}{2}}   \left(\del_\mu
\varphi(x)\right)^2  + \frac{r}{2} \, \varphi^2(x) + \frac{u}{4!}
\,\varphi^4(x) \right\rbrace \,. \eeq
 The NPRG constructs  a family of effective actions, $\Gamma_\kappa[\phi]$ (with $\phi$ the expectation value of the field in the presence of external sources),
in which  the magnitude of long wavelength fluctuations are
controlled by an infrared regulator   depending on a continuous
parameter $\kappa$. The effective action $\Gamma_\kappa[\phi]$
interpolates between the classical action obtained for
$\kappa=\Lambda$ (with $\Lambda$ the microscopic scale at which
fluctuations are essentially suppressed), and the full effective
action obtained when  $\kappa \to 0$, i.e., when all fluctuations
are taken into account (see e.g. \cite{Berges02}). It is understood
that the values of the parameters $r$ and $u$ of the classical
action (\ref{eactON}), as well as the field normalisation, are fixed
at the microscopic scale $\Lambda$. One can write for
$\Gamma_\kappa[\phi ]$ an exact flow equation
\cite{Tetradis94,Morris94,Morris94c}: \beq \label{NPRGeq}
\partial_\kappa \Gamma_\kappa[\phi]=\frac{1}{2} \int \frac{d^dq}{(2\pi)^d}
(\partial_\kappa R_\kappa(q))
\left[\Gamma_\kappa^{(2)}+R_\kappa\right]^{-1}_{q,-q},
\eeq
where $\Gamma_\kappa^{(2)}$ is the second derivative of
$\Gamma_\kappa$ with respect to $\phi$,
and $R_\kappa$ denotes a family of ``cut-off functions''
depending on $\kappa$. There is a large freedom in the choice of
$R_\kappa(q)$, abundantly discussed in the literature
\cite{Ball95,Comellas98,Litim,Canet02}. To be specific,  in the present paper, we shall use for $R_\kappa(q)$
the following function \cite{Litim}
\beq \label{reg-litim}
R_\kappa(q) =Z_\kappa (\kappa^2-q^2) \; \Theta(\kappa^2-q^2),
\eeq
where $Z_\kappa$ is a function of $\kappa$ specified in the next section (see eq.~(\ref{def-Zk})).

By deriving eq.~(\ref{NPRGeq}) with respect to $\phi$, and then
letting the field be constant, one gets the flow equation for the
$n$-point functions $\Gamma^{(n)}$ in a constant background field
$\phi$. More precisely, taking into account momentum conservation,
one defines: \beq\label{gamman} &&(2\pi)^d
\;\delta^{(d)}\left(p_1+\cdots
+p_n\right)\;\Gamma_\kappa^{(n)}(p_1,\dots,p_n;\phi)\qquad\qquad\nonumber
\\&& \qquad\qquad   =\int d^dx_1\dots\int d^dx_{n} e^{i\sum_{j=1}^n
p_jx_j}\left. \frac{\delta^n\Gamma_\kappa}{\delta\phi(x_1) \dots
\delta\phi(x_n)}\right|_{\phi(x)\equiv \phi}. \eeq Then, the
equation for the 2-point function reads:
\begin{eqnarray}
\label{gamma2champnonnul}
\partial_\kappa\Gamma_{\kappa}^{(2)}(p;\phi)&=&\int
\frac{d^dq}{(2\pi)^d}(\partial_\kappa R_\kappa(q))\left\{G_{\kappa}(q;\phi)
\Gamma_{\kappa}^{(3)}(p,q,-p-q;\phi)\right. \nonumber \\
&&\qquad\qquad\qquad\qquad\times G_{\kappa}(q+p;\phi)\Gamma_{\kappa}^{(3)}(-p,p+q,-q;\phi)
G_{\kappa}(q;\phi) \nonumber
 \\
&&\qquad\qquad\qquad\qquad\left.-\frac{1}{2}G_{\kappa}(q;\phi)\Gamma_{\kappa}^{(4)}
(p,-p,q,-q;\phi)G_{\kappa}(q;\phi)\right\} ,
\end{eqnarray}
where
\begin{equation}\label{G-gamma2}
G^{-1}_{\kappa} (q;\phi) \equiv \Gamma^{(2)}_{\kappa} (q,-q;\phi) +
R_\kappa(q),
\end{equation}
and in eq.~(\ref{gamma2champnonnul}) we have used the simplified
notation $\Gamma^{(2)}_{\kappa} (q;\phi)$ for $\Gamma^{(2)}_{\kappa}
(q,-q;\phi)$, a notation that will be used throughout.

In general, the flow equation for a given $n$-point function
involves the $m$-point functions with $m=n+1$ and $m=n+2$. Thus, the
flow equations for the $n$-point functions do not close, but
constitute an infinite hierarchy of coupled equations; this makes
them difficult to solve.

In \cite{Blaizot:2005xy} we proposed a method to solve this infinite
hierarchy. It exploits the smoothness of the regularized $n$-point
functions at small momenta, and the fact that the loop momentum $q$
in the right hand side of the flow equations (such as
eq.~(\ref{NPRGeq}) or eq.~(\ref{gamma2champnonnul})) is limited to
$q\simle \kappa$ by the presence of the regulator $R_\kappa(q)$. The
leading order (LO) of the method presented in \cite{Blaizot:2005xy}
thus consists in setting $q=0$ in the $n$-point functions in the
r.h.s. of the flow equations, for instance \beq
\Gamma^{(n)}_{\kappa}(p_1,p_2,...,p_{n-1}+q,p_n-q)\sim
\Gamma^{(n)}_{\kappa}(p_1,p_2,...,p_{n-1},p_n). \eeq  Once this
approximation is made, some momenta in some of the $n$-point
functions vanish, and the corresponding $n$-point functions can be
obtained as the derivatives of  $m$-point functions ($m<n$) with
respect to a constant background field, thereby allowing us to close
the hierarchy of equations.

 Specifically,
in eq.~(\ref{gamma2champnonnul}) for the 2-point function, the 3-
and 4-point functions in the r.h.s. will contain respectively one
and two vanishing momenta after we set $q=0$. These  can be related
to the following derivatives of the 2-point function:
\beq\label{derivs}
\Gamma_{\kappa}^{(3)}(p,-p,0;\phi)=\frac{\partial
\Gamma_{\kappa}^{(2)} (p;\phi)} {\partial \phi} , \hskip 1 cm
\Gamma_{\kappa}^{(4)}(p,-p,0,0;\phi)=\frac{\partial^2
\Gamma_{\kappa}^{(2)} (p;\phi)} {\partial \phi^2}.
\eeq
One then arrives at a closed equation for
$\Gamma_{\kappa}^{(2)}(p;\rho)$ (with $\rho\equiv \phi^2/2$):
\beq
 \label{2pointcloseda}
\kappa \partial_\kappa\Gamma_\kappa^{(2)}(p;\rho)=
J_d^{(3)}(p;\kappa;\rho) \; \left( \frac{\partial
\Gamma_\kappa^{(2)}(p;\rho)} {\partial \phi} \right)^2
  -\frac{1}{2} I_d^{(2)}(\kappa;\rho) \; \frac{\partial^2
\Gamma_\kappa^{(2)}(p;\rho)} {\partial \phi^2},
\eeq
where \beq\label{defJ} J_d^{(n)}(p;\kappa;\rho)\equiv
\int\frac{d^dq}{(2\pi)^d}\kappa (\partial_\kappa R_\kappa(q))
G_\kappa(p+q;\rho)G^{(n-1)}_\kappa(q;\rho) , \eeq and
\beq\label{defI} I_d^{(n)}(\kappa;\rho)\equiv \int
\frac{d^dq}{(2\pi)^d}\kappa (\partial_\kappa R_\kappa(q))
G^n_\kappa(q;\rho). \eeq  Note that
$J_d^{(n)}(p=0;\kappa;\rho)=I_d^{(n)}(\kappa;\rho)$.

At this point we note that the $n$-point functions at zero external momenta can all be considered as derivatives of a single function, the effective potential $V_\kappa(\rho)$. For instance, 
 \beq
\label{zeromomenta} \Gamma^{(2)}_\kappa (p=0;\rho)=\frac{\partial^2
V_\kappa}{\partial \phi^2}.
\eeq  The effective potential 
satisfies a  flow equation which can be deduced  from that for the effective action, eq.~(\ref{NPRGeq}), when restricted to constant fields. It reads \beq\label{eqforV} \kappa\partial_\kappa
V_\kappa(\rho)=\frac{1}{2}\int \frac{d^dq}{(2\pi)^d} \kappa
(\partial_\kappa R_\kappa(q)) G_\kappa(q;\rho). \eeq The second  derivative  of this equation with respect to the background field yields a flow equation for 
$\Gamma^{(2)}_\kappa (p=0;\rho)$. Now, this  equation  does not coincide with eq.~(\ref{2pointcloseda}) in
which we set $p=0$: indeed, in contrast to eq.~(\ref{2pointcloseda}), the vertices in the equation deduced from eq.~(\ref{eqforV}) keep their $q$-dependence ($q$ being the loop momentum in eq.~(\ref{eqforV})). There is therefore an apparent inconsistency in our approximation scheme, that is  however easily resolved by treating separately the zero momentum ($p=0$) and the non-zero momentum ($p\ne 0$) sectors.   In fact, in doing so,  we  get more accuracy in the sector $p=0$ than in the sector  $p\ne 0$.

Let us then write : 
\beq\label{defsigmabis}
\Gamma^{(2)}_\kappa (p;\rho)=   p^2 +\frac{\partial^2
V_\kappa}{\partial \phi^2}+\Sigma_\kappa (p;\rho),
 \eeq
 where 
 \beq\label{defsigma} \Sigma_\kappa
(p;\rho) \equiv \Gamma^{(2)}_\kappa (p;\rho) - p^2 -
\Gamma^{(2)}_\kappa (p=0;\rho). \eeq 
We shall refer to $\Sigma_\kappa (p;\rho)$ as the self-energy (although it differs from the usual self-energy by the subtraction of the momentum independent  contribution $\Gamma^{(2)}_\kappa (p=0;\rho)$). By definition, $\Sigma_\kappa (p=0;\rho)=0$, and at criticality, 
$\Gamma^{(2)}_{\kappa=0} (p=0;\rho)=0$. We then  proceed  with separate approximations in the two sectors with $p=0$ and $p\ne 0$. 

In the sector  $p\ne 0$, it is $\Sigma_\kappa (p;\rho)$, rather than $\Gamma^{(2)}_\kappa (p;\rho)$ which satisfies the approximate eq.~(\ref{2pointcloseda})
(strictly speaking, eq.~(\ref{2pointcloseda}) to which one subtracts
the same equation with  $p=0$):
\beq
\label{2pointclosed}
\kappa \partial_\kappa\Sigma(p;\rho)=\left[
J_d^{(3)}(p;\kappa;\rho) \; \left( \frac{\partial
\Gamma_\kappa^{(2)}(p;\rho)} {\partial \phi} \right)^2
  -\frac{1}{2} I_d^{(2)}(\kappa;\rho) \; \frac{\partial^2
\Gamma_\kappa^{(2)}(p;\rho)} {\partial \phi^2}\right]-[p\to 0].
\eeq
Eq.~(\ref{2pointclosed}) is the flow equation for the momentum dependent part of the 2-point
function at LO of our approximation scheme.  It is a partial
differential equation with respect to the two real variables,
$\kappa$ and $\rho$, with the momentum $p$ playing the role of a
parameter. It is to be integrated from $\kappa=\Lambda$, with
 initial condition  $\Sigma_\Lambda (p;\rho)=0$ 
 (see eqs.~(\ref{eactON}) and (\ref{defsigma})),  to
$\kappa=0$ where it yields the physical self-energy $\Sigma (p;\rho) \equiv \Sigma_{\kappa=0}
(p;\rho) $.

Eq.~(\ref{2pointclosed}) is to be solved together with the equation in the sector $p=0$, i.e., with the equation for   the effective potential, with initial condition $V_\Lambda(\rho)= r \rho+ (u/6) \rho^2$ (see eq.~(\ref{eactON})). In  eq.~(\ref{eqforV}) for $V_\kappa(\rho)$, we use the propagator (\ref{defsigmabis}) in which  $\Sigma_\kappa (p;\rho)$ is solution of eq.~(\ref{2pointclosed}) and   $\Gamma^{(2)}_\kappa (q=0;\rho)$ is determined self-consistently from the effective potential, 
using eq.~(\ref{zeromomenta}). 

It is not difficult to verify that (in the perturbative regime) this scheme has 2-loop accuracy for the effective potential, and only one-loop accuracy for the self-energy. Besides, in the low momentum region it is as accurate as the derivative expansion at next-to-leading order.

\section{Analysis of the flow equation}

\label{analysis}

There are two features of  eq.~(\ref{2pointclosed}) that make it  a priori difficult to solve.
 First, the two functions $J_d^{(3)}(p;\kappa;\rho)$
and $I_d^{(2)}(\kappa;\rho)$,  are functionals of  the solution
$\Gamma^{(2)}_\kappa(p;\rho)$ (see eq.~(\ref{G-gamma2})). Second the
different values of $p$ are coupled through the propagator
$G_\kappa(p+q)$ entering the calculation  of
$J^{(3)}_d(p;\kappa;\rho)$. In principle, one should therefore solve
eq.~(\ref{2pointclosed}) self-consistently, and simultaneously for
all values of $p$. However, in this section, we shall show that it
is possible to make an accurate calculation of $J^{(3)}_d(p;\kappa
;\rho)$ and $I_d^{(2)}(\kappa;\rho)$ using approximate propagators.
This yields an approximate version of eq.~(\ref{2pointclosed}) that
can be solved for each  given value of $p$. The validity of this
approximation will be checked in the next section.

Consider first
the function $I_d^{(n)}(\kappa;\rho)$, which does not depend on $p$.
The smoothness of the $n$-point functions and the fact that $q\leq
\kappa$,  suggest to perform in the propagators of the
right-hand-side of eq.~(\ref{defI}) an approximation similar to that
applied to the other $n$-point functions, i.e., set $q=0$. However,  in order to maintain the exact one-loop properties
of the flow equations,  one cannot simply set $q=0$ in the propagators:  rather, one needs to keep a momentum dependence close to that of the free propagators. Thus, we shall  use for  the propagators entering the
calculation of $I^{(n)}_d(\kappa;\rho)$ the following approximate
form \beq\label{propGq} G_\kappa^{-1}(q;\rho) \approx Z_\kappa q^2 +
\Gamma_\kappa^{(2)}(q=0;\rho) + R_\kappa(q), \eeq where \beq
\label{def-Zk} Z_\kappa \equiv\left. {\frac{\partial
\Gamma_\kappa^{(2)}}{\partial q^2}}\right|_{q=0,\rho=\rho_0}. \eeq
As well known \cite{Morris94c}, and will be verified in app.~A,  $\left.\partial \Gamma^{(2)}(q;\rho)/\partial q^2\right|_{q=0}$ depends weakly on $\rho$. Accordingly, one expects $Z_\kappa$ to depend weakly on  the value chosen for $\rho_0$. As will be seen  in app. A,  the choice 
 $\rho_0=0$ is here the simplest. With the propagator
(\ref{propGq}), and the function (\ref{reg-litim}) for $R_\kappa(q)$
one can calculate   $I_d^{(n)}(\kappa;\rho)$ analytically: \beq
\label{In-anal} I_d^{(n)}(\kappa;\rho) = 2 K_d
\frac{\kappa^{d+2-2n}}{Z_\kappa^{n-1}} \frac{1}{(1+\hat
m^2_\kappa(\rho))^n} \left(1-\frac{\eta_\kappa}{d+2}\right). \eeq In
this expression, \beq\label{defZk}
\eta_\kappa\equiv-\kappa\partial_\kappa \ln Z_\kappa \eeq is the
running anomalous dimension and \beq\label{defm2k} \hat
m^2_\kappa(\rho) \equiv \frac{\Gamma_\kappa^{(2)}(q=0;\rho) }{ \kappa^2
Z_\kappa}, \eeq is a dimensionless, field-dependent, effective mass.
$K_d$ is a number resulting from angular integration,
$K_d^{-1}\equiv d\; 2^{d-1}\; \pi^{d/2} \; \Gamma(d/2)$ (e.g.,
$K_3=1/(6\pi^2)$). Notice that, for $d>2$, $I_d^{(2)}(\kappa;\rho)
\to 0$ when $\kappa \to 0$.

We shall calculate the function $J^{(3)}_d(p,\kappa;\rho)$ in a
similar way, arguing that in this calculation one can assume
$p\simle\kappa$ :  the propagator $G_\kappa(p+q;\rho)$ in
eq.~(\ref{defJ}) is small as soon as $p/\kappa$ is large, and one
can indeed verify  that the function $J^{(3)}_d(p;\kappa;\rho)$
vanishes approximately as $\kappa^2/p^2$ for large values of
$p/\kappa$ (see the explicit expressions (\ref{Jlarge}) and (\ref{Jlarge2}) 
given in app.~B). Thus, in
the region where $J^{(3)}_d(p;\kappa;\rho)$ has a significant value,
one can use for $G_\kappa(p+q;\rho)$ an expression  similar to
(\ref{propGq}), namely   \beq \label{propGpq}
G_\kappa^{-1}(p+q;\rho) \approx Z_\kappa (p+q)^2 +
\Gamma^{(2)}_\kappa(0;\rho) + R_\kappa(p+q). \eeq One can then
calculate the function $J^{(3)}_d(p;\kappa;\rho)$ analytically (in $d=3$). The
resulting expression is more complicated than that of
$I_d^{(2)}(\kappa;\rho)$, eq.~(\ref{In-anal}). It is given in
app.~\ref{functions}  (see also \cite{Blaizot:2006vr}).
 Observe that the regulator in eq.~(\ref{reg-litim}) is
not analytic at $q\sim \kappa$. This generates non analyticities
in $J^{(3)}_d(p;\kappa;\rho)$; but
 these occur  only in the third derivative with respect
  to $p$, at $p=0$ and at $p=2\kappa$ (cf.  the odd powers of $\bar p$ in eqs.~(\ref{Jlarge}-\ref{largeJ4})), and they  play no role at the present level of approximation.

With the approximations just discussed, $I_d^{(2)}(\kappa;\rho)$ and
$J^{(3)}_d(p;\kappa;\rho)$ depend only on quantities that enter the
flow equations at $p=0$, namely $\hat m^2_\kappa(\rho)$ and 
$Z_\kappa$ (or $\eta_\kappa$). 
As we discuss in app.~\ref{LPA}, these quantities
can be obtained from a modified version of the Local Potential
approximation \cite{Berges02} that we call the LPA'. The
strategy to solve eq.~(\ref{2pointclosed}) consists then in two
steps: one first solves the LPA' to get $\hat m^2_\kappa(\rho)$ 
and $\eta_\kappa$; then, for each value of $p$, one
solves eq.~(\ref{2pointclosed}) with the kernels $I_d^{(2)}
(\kappa;\rho)$ and $J_d^{(3)}(p;\kappa;\rho)$ that are calculated
with $\hat m^2_\kappa(\rho)$, $Z_\kappa$ and $\eta_\kappa$
determined from the LPA'.

Note that,  generally, the flow of $\Sigma$ gets strongly 
suppressed  below 
some non vanishing value of $\kappa$. In $d=3$, this can be inferred
from  the properties of the functions $I_{3}^{(2)}(\kappa;\rho)$ and
$J_{3}^{(3)}(p;\kappa;\rho)$ discussed above, and it will be
verified explicitly on the numerical results presented in the next
section (see Fig. \ref{gammaUV}). 
In fact, the  flow of $\Sigma$ receives two
contributions:  the first involves the external momentum $p \neq 0$
and is suppressed when  $\kappa \simle p$ 
($J_{3}^{(3)}(p;\kappa;\rho)$ vanishes rapidly when $\kappa$ 
becomes smaller than $p$,
while $I_{3}^{(2)}(\kappa;\rho) \sim \kappa$); 
the other contribution is independent
of $p$ and, at criticality,  is suppressed for $\kappa \simle \kappa_c \sim u/10$ (see app.~A, and in particular fig.~\ref{m2k}). Accordingly,
one expects the flow to stop when $\kappa$ reaches the smallest of
$\kappa_c$ and $p$.

 The function $\Gamma_\kappa^{(2)}(p;\rho)$  exhibits a simple scaling
behavior.  Consider for simplicity the zero field case $\rho=0$, and
the ratio
\begin{equation}
\frac{p^2+\Gamma_\kappa^{(2)}(0;\rho=0)+\Sigma_\kappa(p;\rho=0)}{\Gamma_
\kappa^{(2)}(0;\rho=0)}=f\left(\frac{p}{\kappa},\frac{p}{u}\right).
\end{equation}
At criticality, and  in the scaling regime where $p,\kappa\ll u$, we
expect $f$ to become independent of $u$, and therefore a function
of $p/\kappa$ only. As will be shown in the next section, the
solution of eq.~(\ref{2pointclosed}) verifies this property.
Note that this scaling behavior is reproduced only
when including a renormalization factor $Z_\kappa$ whose flow is
determined consistently from  that of ${\partial
\Sigma_\kappa}/{\partial p^2}$ for $p< \kappa$, as obtained
from eq.~(\ref{2pointclosed}). This calculation of $Z_\kappa$ is
explained in app.~A. We have tested the consequence of setting $Z_\kappa=1$ in the propagators (\ref{propGq}) and (\ref{propGpq}), corresponding to the Local Potential approximation (as opposed to the LPA'). Doing so does not alter the self-energy in any significant way when $p\simge u$, but in the IR regime, the scaling behavior is only approximate.  
 
\section{Numerical results and discussion}
\label{numerics}

We now turn to the numerical solution of the flow equation for
$\Sigma_\kappa (p;\rho)$, at $d=3$ and at criticality. Our goal is
to assess the quality of the approximation scheme, and there are two
aspects that we shall examine.  First, since the strategy described
in the previous section provides only an approximate solution to
eq.~(\ref{2pointclosed}), we shall  estimate by how much this
approximate solution differs from the exact solution of this
equation. Second, since eq.~(\ref{2pointclosed}) itself is only the
LO approximation of the method described in \cite{Blaizot:2005xy},
we shall compare our results with known ones concerning the
self-energy of the scalar model at criticality.

\begin{figure}[t]
\begin{center}
\includegraphics*[scale=0.28,angle=-90]{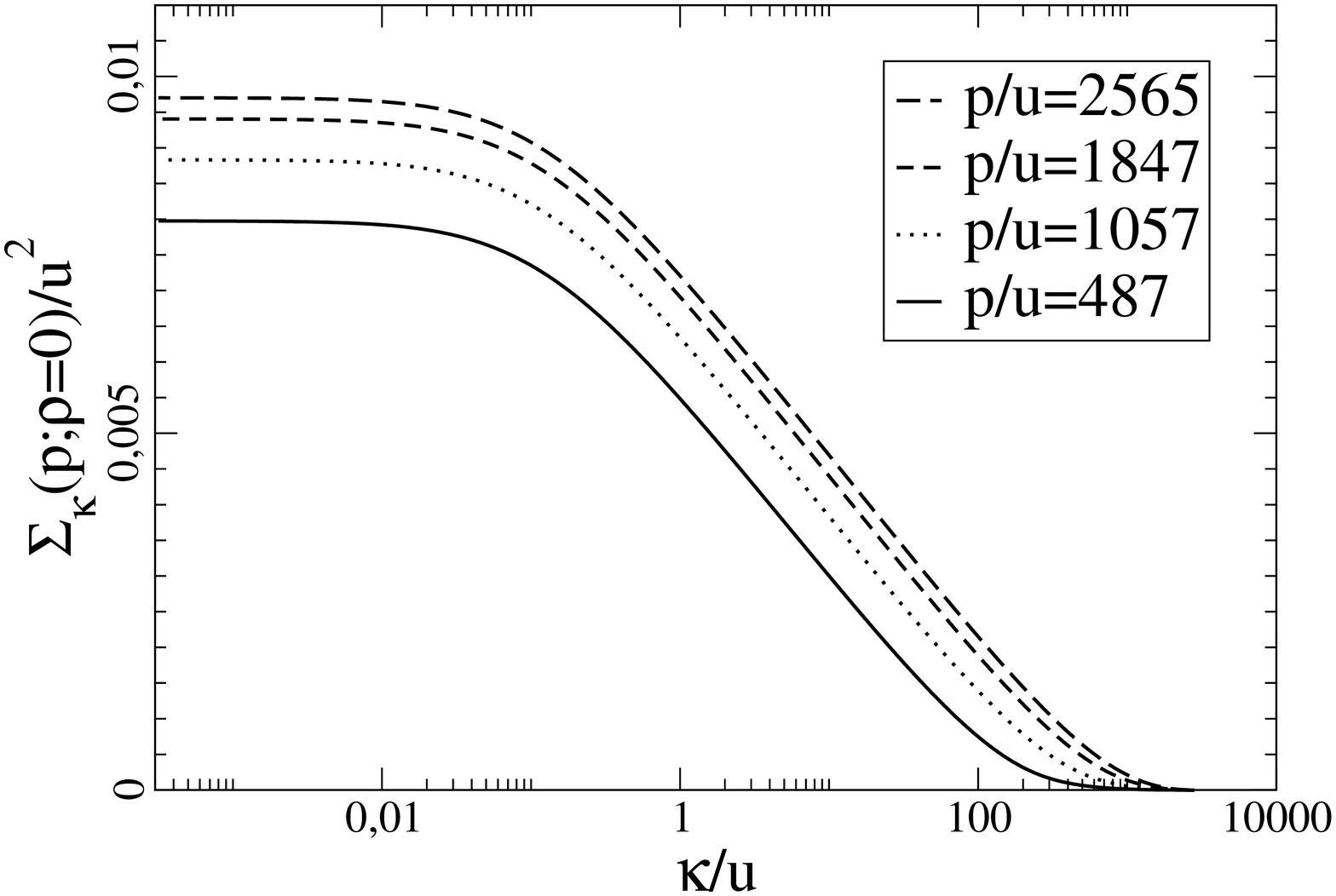}
\includegraphics*[scale=0.28,angle=-90]{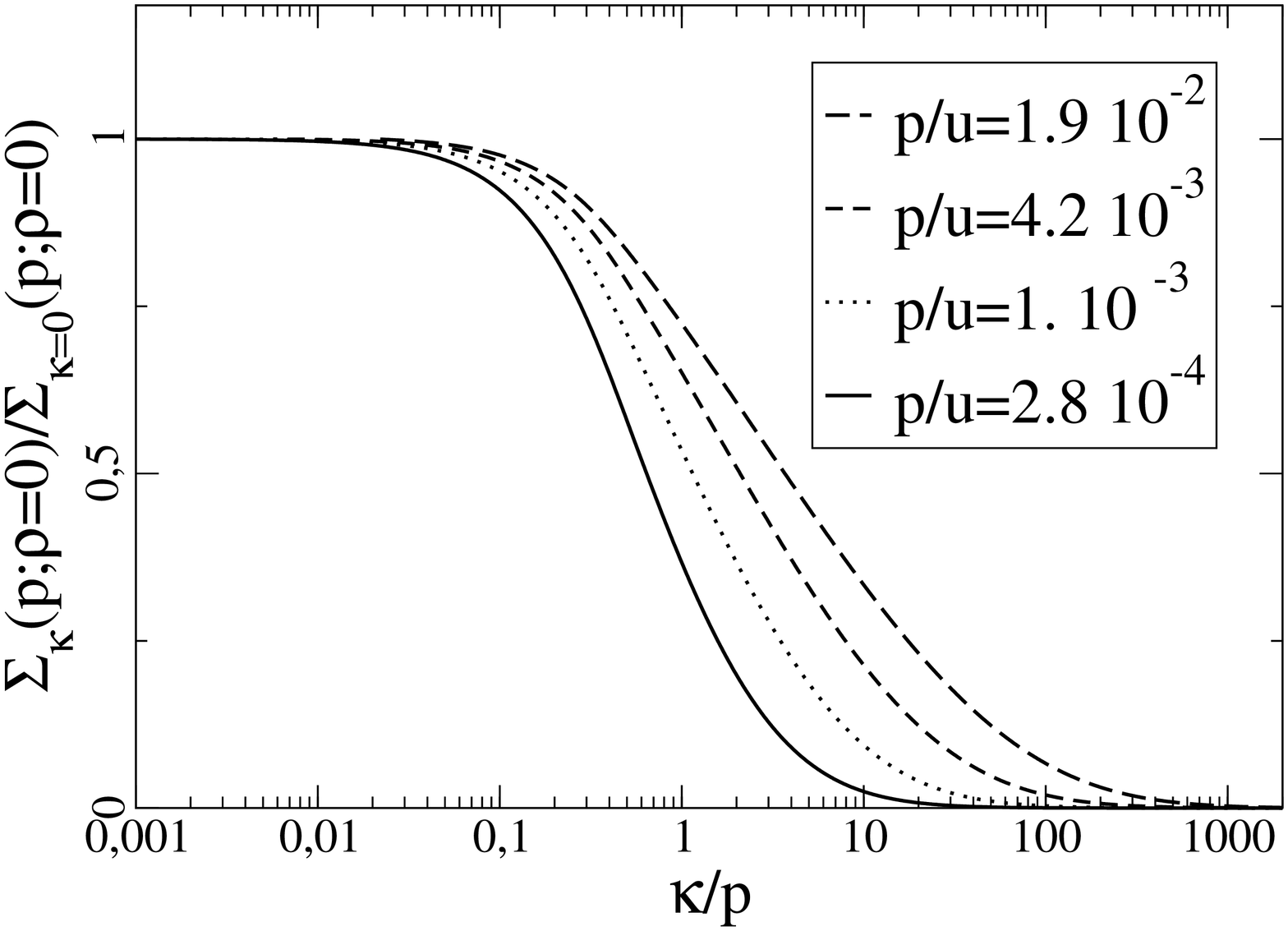}
\end{center}
\caption{ \label{gammaUV} Left: $\Sigma_\kappa(p;\rho=0)/u^2$
 as a
 function of
$\kappa/u$  for various values of $p\gg u$. The flow stops at
$\kappa \simle \kappa_c \sim u/10$. Right:
$\Sigma_\kappa(p;\rho=0)/\Sigma_{\kappa=0}(p;\rho=0)$ as a function of
$\kappa/p$ for various values of $p\ll u$. The flow stops at $\kappa
\simle p/10$. }
\end{figure}

Let us start by considering general properties of the flow, and
verify in particular that it essentially stops for a small value 
of $\kappa$. Fig.~\ref{gammaUV}
displays the self-energy, $\Sigma_\kappa (p;\rho=0)$ as a function
of $\kappa/u$, for different values of $p$. Calculations  are made
for $u/\Lambda=3.54\times 10^{-4}$  (this value is small enough to guarantee
that the results are independent of $\Lambda$).  The left panel of
fig.~\ref{gammaUV} shows the flow of $\Sigma_\kappa (p;\rho=0)$ for
values of $p$ in the UV regime, i.e., $p \gg \kappa_c\sim u/10$; for
all the considered values of $p$ the flow stops at
 $\kappa_c $. The right panel of fig.~\ref{gammaUV} presents the flow of the self-energy  when $p$ is in the
IR regime, i.e., when $p \ll \kappa_c$. In this case,  we have
divided $\Sigma_\kappa (p;\rho=0)$ by its physical value 
$\Sigma(p;\rho=0)\equiv \Sigma_{\kappa=0} (p;\rho=0)$, in order to make it more obvious
that the  flow only stops when $\kappa\simle p/10$.

\begin{figure}[t]
\begin{center}
\includegraphics*[scale=0.4,angle=-90]{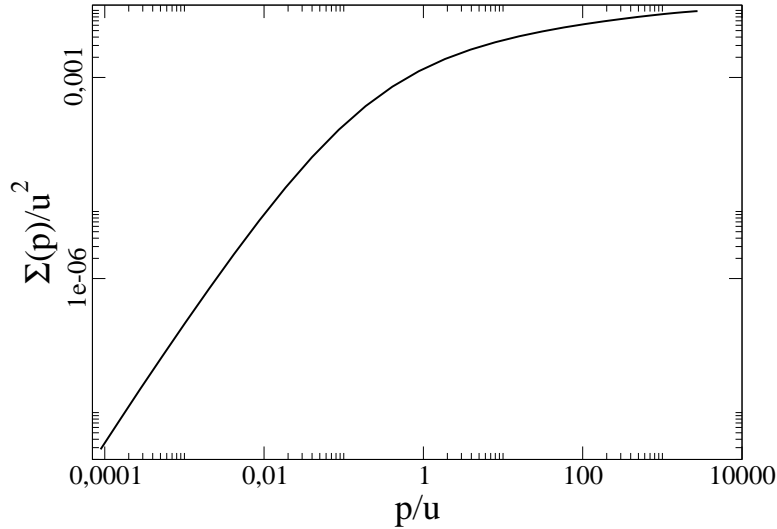}
\end{center}
\caption{ \label{self}  $\Sigma(p)/u^2$, in $d=3$, at
criticality and zero external field, as a function of $p/u$. }
\end{figure}

We now turn to the physical  self-energy  $\Sigma(p) \equiv
\Sigma_{\kappa=0}(p;\rho=0)$ in vanishing external field, displayed in  fig.~\ref{self} as a
function of $p/u$, and discuss its behavior in the various momentum
regions: $p\gg u$, $p\ll u$, $p\sim \kappa_c\sim u/10$. We have checked that the curve in  fig.~\ref{self}, i.e., $(1/u^2) \Sigma(p/u)$,  is ``universal'', i.e., independent of $u$ and $\Lambda$, provided $u/\Lambda$ is small enough.

In the perturbative regime ($p\gg u$), one expects $\Sigma(p)\approx
(u^2/96\pi^2)\log(p/u)$. In app.~\ref{perturbative} we show that the
analytical solution of eq.~(\ref{2pointclosed}) preserves this
behavior, although the coefficient in front of the logarithm is $(u^2/9\pi^4)$, 
 8\% larger (the LO approximation does not  include all the
2-loop perturbative diagrams  exactly). Our approximate numerical
solution reproduces this result. As
explained in \cite{Blaizot:2005xy}, at the NLO of our approximation
scheme, which is beyond the scope of the present paper,  the
contribution of the 2-loops diagrams would be  exactly included and
the correct prefactor $(u^2/96\pi^2)$ would be recovered.

 In the IR
region ($p\ll u$)  we expect the self-energy to behave as \beq
p^2+\Sigma(p) = A p^{2-\eta^*}, \eeq where $\eta^*$ is the anomalous
dimension.  By analyzing the small momentum  behavior of
$\Sigma(p)$, we get numerically $\eta^* = 0.05218$.   An alternative
way to determine $\eta^*$ is to extract it from the $\kappa$
dependence of $Z_\kappa$ (see eq.~(\ref{defZk})). As recalled in
app.~A, in the critical regime, i.e., when $\kappa\simle\kappa_c$,
$Z_\kappa \propto \kappa^{-\eta^*}$, with $\eta^*=0.05220$ the fixed point value of $\eta_\kappa$ (see fig.~\ref{m2k}). It is also shown in fig.~\ref{m2k} in app.~A  that the quantity $\Gamma^{(2)}_\kappa(p=0;\rho=0)/(\kappa^2 Z_\kappa)$ goes to a fixed point, which confirms the behavior of $\Gamma^{(2)}_\kappa(p=0;\rho=0)\sim \kappa^{2-\eta}$ expected in the scaling regime.

\begin{figure}[t]
\begin{center}
\includegraphics*[scale=0.28,angle=-90]{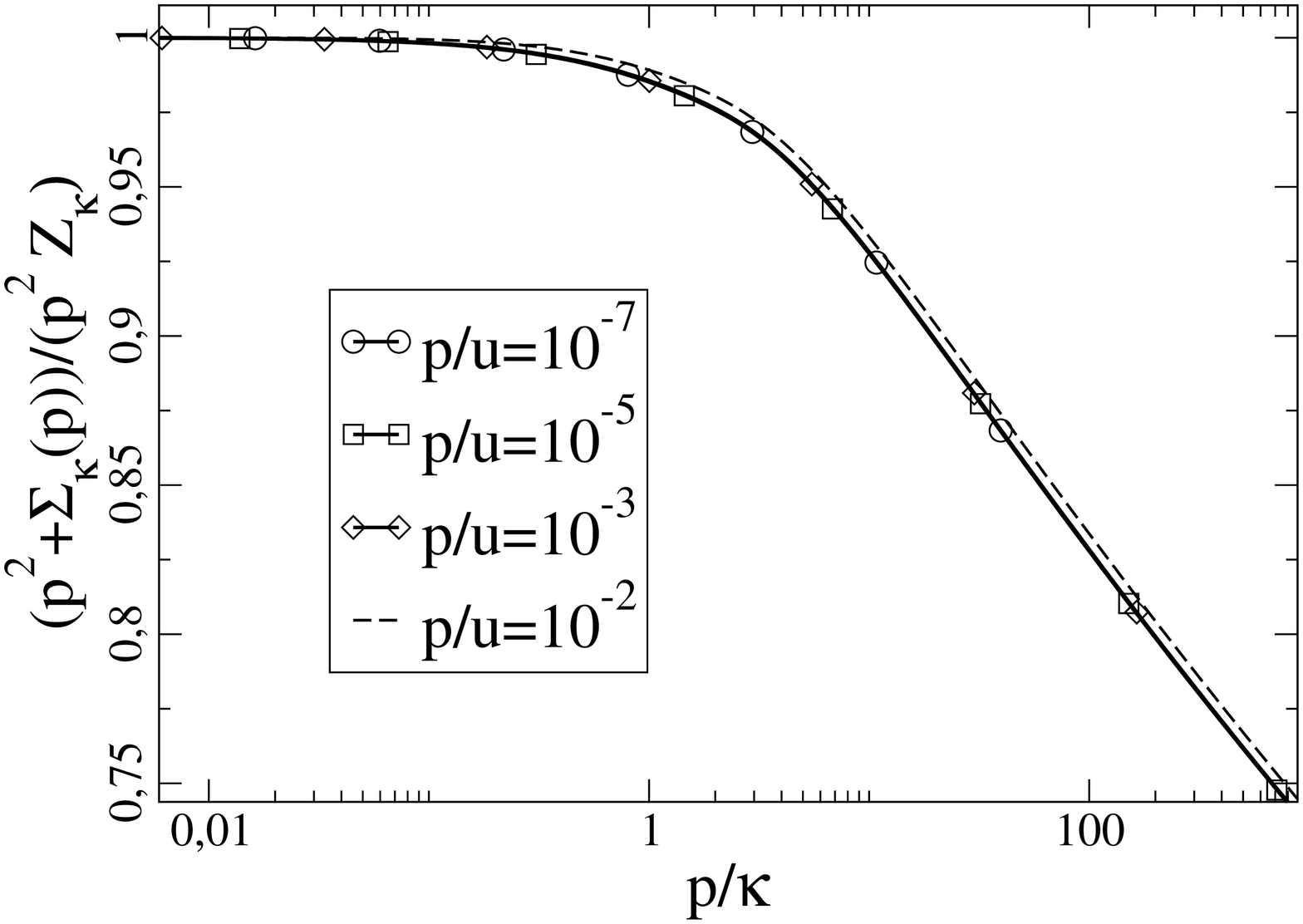}
\includegraphics*[scale=0.28,angle=-90]{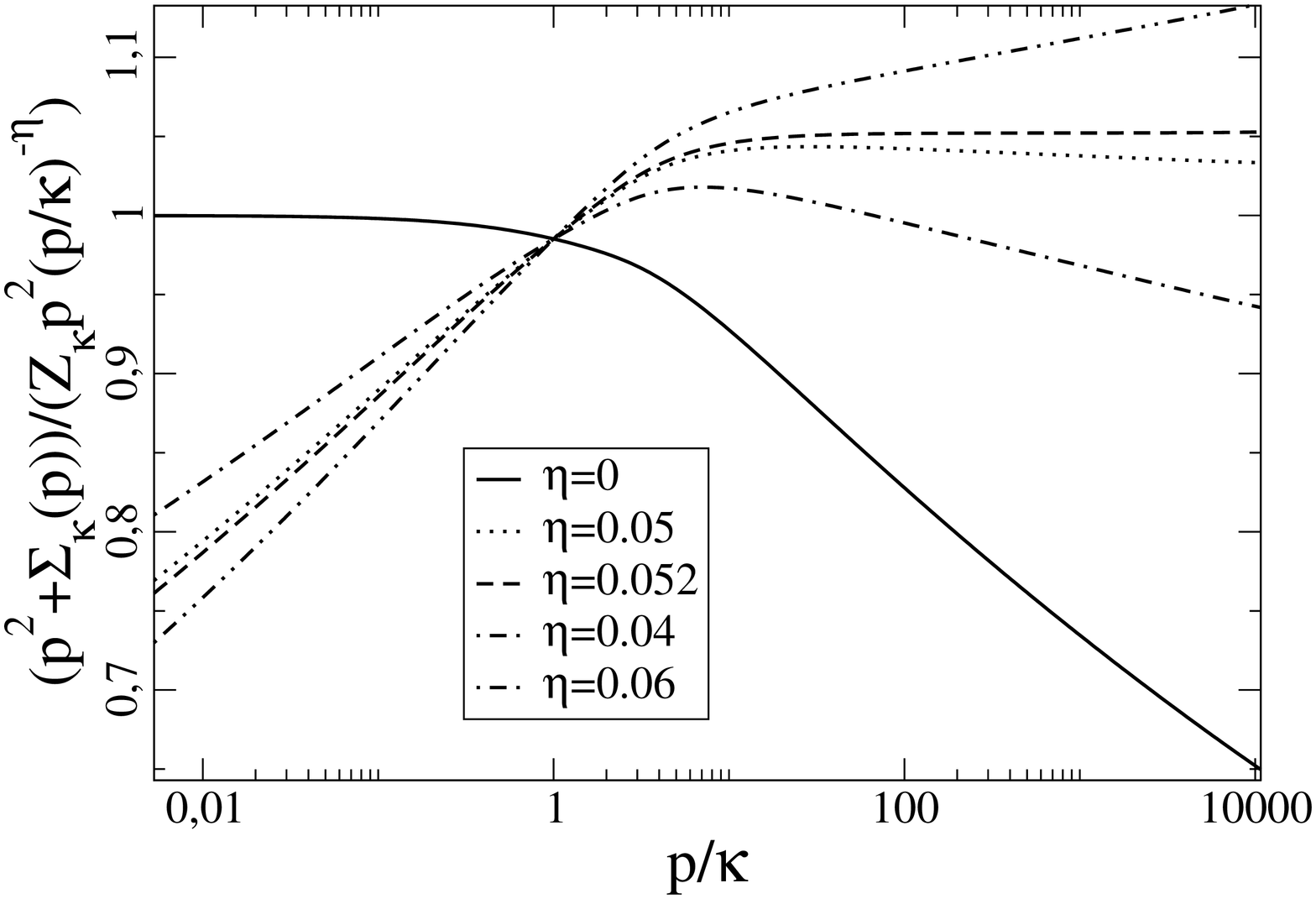}
\end{center}
\caption{ \label{scaling1} The ratio
${(p^2+\Sigma_\kappa(p))}/({Z_\kappa p^2})$ as a functio of
$p/\kappa$. Right: the same ratio divided by $(p/\kappa)^{-\eta^*}$.}
\end{figure}

We have performed a more stringent test of scaling by studying
the function $ {(p^2+\Sigma_\kappa(p))}/({Z_\kappa p^2})$. This
function is displayed in fig.~\ref{scaling1} as a function of
$p/\kappa$. By definition of $Z_\kappa$ (see eq.~(\ref{def-Zk})),
when $\kappa$ is kept fixed and $p\to 0$, this function goes to one.
Furthermore, as explained before, in the scaling regime $p,\kappa
\ll u$, one expects this function to depend on $p/\kappa$ only,
which is indeed well verified, as can be seen on the left panel of
fig.~\ref{scaling1}; it is only for values of $p$ which are not
small enough ($p/u=10^{-2}$, corresponding to the dashed line)  that
 violation of this scaling starts to become significant. Moreover, 
 as can be seen on the figure, $p^2+\Sigma_\kappa(p)$ is  well 
 approximated by $Z_\kappa p^2$ for all $p\simle \kappa$. In the 
 right panel of
fig.~\ref{scaling1},  we have plotted the ratio  $
{(p^2+\Sigma_\kappa(p))}/({Z_\kappa p^2})$  divided by
$(p/\kappa)^{-\eta^*}$. Recall that when $\kappa\ll p\ll \kappa_c$,
one expects $p^2+\Sigma_\kappa(p)\sim p^{2-\eta^*}$, while
$Z_\kappa\sim \kappa^{-\eta^*}$ when $\kappa\simle \kappa_c$.
Therefore when $1\ll p/\kappa\ll\kappa_c/\kappa$, one expects
${(p^2+\Sigma_\kappa(p))}/({Z_\kappa p^2})\sim
(p/\kappa)^{-\eta^*}$, so that the quantity which is plotted should
be constant. As seen in the right panel of fig.~\ref{scaling1}, this
is indeed the case for the value $\eta^*=0.05219$, which confirms the coherence of the whole calculation.

Our  estimate for the anomalous dimension, $\eta^*\approx 0.052$ is to be compared with
the results $\eta^*=0$, $\eta^*=0.044$ and $\eta^*=0.033$ obtained
with the derivative expansion at LO, NLO and NNLO, respectively
\cite{Morris94c,Canet02,Canet03}, and $\eta^* = 0.0335 \pm 0.0025$ from the resummed 7
loop calculation of ref.~ \cite{refeta}. Thus, the LO of our approximation scheme
yields a result slightly larger that the NLO of the derivative 
expansion. The value of $\eta^*$ obtained here is also slightly 
larger than that obtained in \cite{Blaizot:2005wd} using a 
different version of the LPA'  than that used here. In fact, the 
deviation of the present estimate of $\eta^*$ from the value 
$0.044$ obtained with the derivative expansion in next-to-leading 
order can be taken as a measure of the error introduced, in the 
scaling regime, by our use 
of the LPA' in our approximate solution of eq.~(\ref
{2pointclosed}): as already mentioned, if we had solved eq.~(\ref
{2pointclosed}) exactly, one should have obtained essentially the 
same value as in the derivative expansion at NLO.

We turn now to the intermediate momentum
region, which we shall probe with a quantity which is very sensitive to
the cross-over  between the two regimes just studied:
\beq\label{integralc} \Delta\langle
\phi^2\rangle= \int\frac{d^3 p}{(2\pi)^3}\,\left(
\frac{1}{p^2+\Sigma(p)}-\frac{1}{p^2}\right).
\eeq
As shown for instance in \cite{Blaizot:2004qa}, the  integrand in eq.~(\ref{integralc})  is peaked at $p\sim\kappa_c$ (in fact it takes  significant values only in the region $10^{-3} \simle p/u \simle 10$). This quantity has been much studied recently  for a scalar model with $O(2)$ symmetry  because it determines then the shift of the critical
temperature of the weakly repulsive Bose
 gas
\cite{club}. For the simple scalar model studied in this paper,  lattice calculations measure
\cite{latt3} $\Delta\langle \phi^2\rangle/u = -(4.95 \pm 0.41)
\times 10^{-4}$ while the ``7 loop" resummed calculation of ref.~\cite{Kastening:2003iu} yields $\Delta\langle \phi^2\rangle/u = -(4.86 \pm 0.45)
\times 10^{-4}$. With the present numerical
 solution, one gets
$\Delta\langle \phi^2\rangle/u =- 5.39 \times 10^{-4}$.   This is only slightly larger than the value
$\Delta\langle \phi^2\rangle/u = -5.03\times 10^{-4}$ obtained in the next-to-leading order of the scheme presented in  \cite{Blaizot:2005wd,Blaizot:2006vr}.

We conclude that with the LO of the present approximation scheme, we
obtain an accurate description of the self-energy in the entire
range of momenta. Since we have solved only appproximately eq.~(\ref{2pointclosed}), it remains to study by  how much the solution that
we have obtained differs from  the exact solution of
eq.~(\ref{2pointclosed}). We have already indications about the accuracy of the approximation both in the UV and in  the IR. In the UV, we reproduce the expected result (which differs by 8\% from the exact 2-loop result). We also loose the 2-loop accuracy with which the efffective potential could be obtained in the LO of our scheme, 
by using LPA' propagators. In the IR, we have seen that the result that we got for the anomalous dimension is 0.052 instead of a  value close to 0.044 that would have been obtained had we solved exactly eq.~(\ref{2pointclosed}). As a further test, we have recalculated $I^{(2)}_{3}(\kappa;\rho)$ and $J_{3}^{(3)}(p;\kappa;\rho)$ using,
instead of the LPA' propagators, the  propagators (\ref{G-gamma2})
in which  $\Gamma^{(2)}_\kappa(p;\rho)$ is the 2-point function that
has been obtained in  this section by approximately solving eq.~(\ref{2pointclosed}).

\begin{figure}[t]
\includegraphics*[scale=0.28,angle=-90]{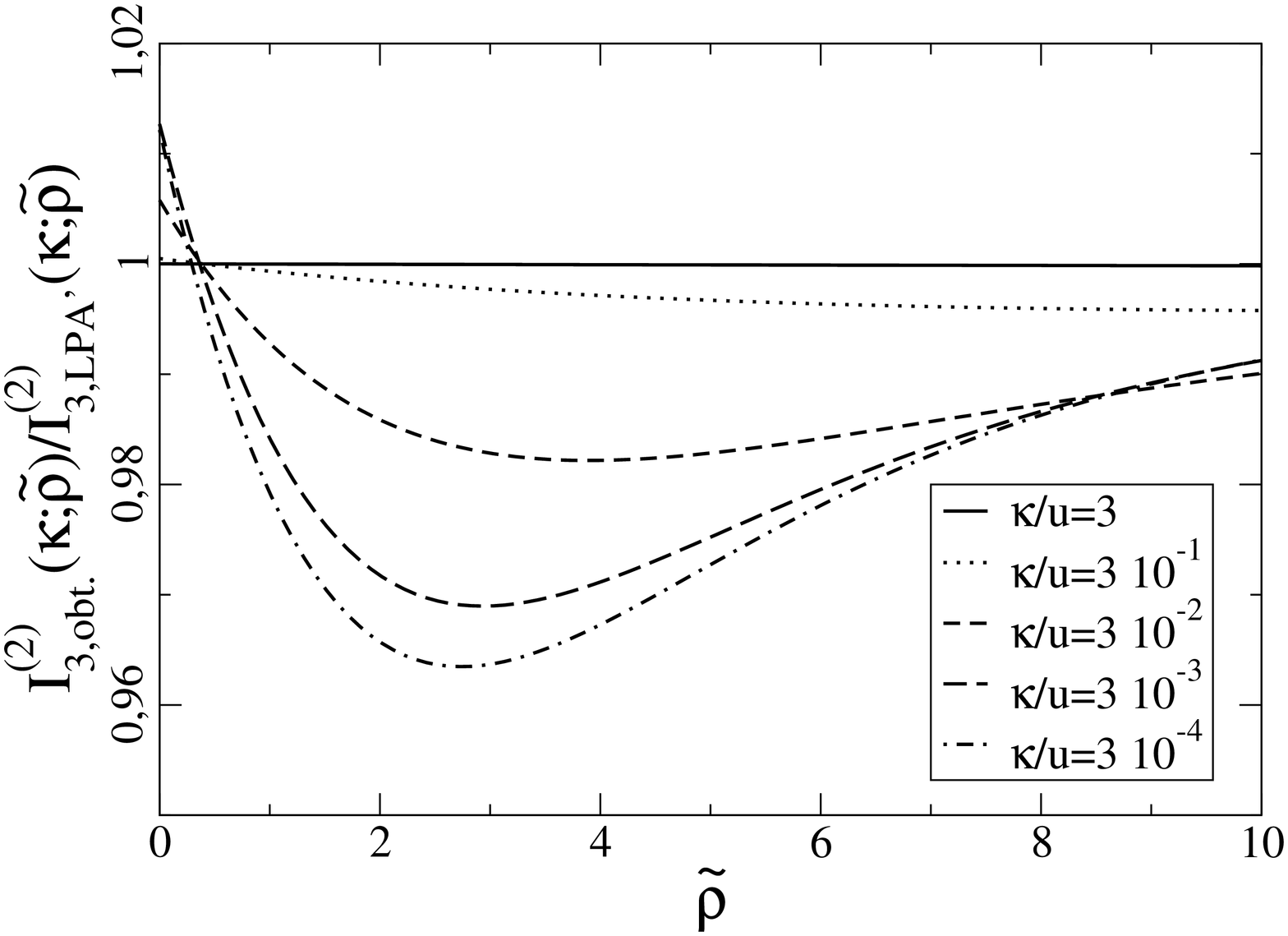}
\includegraphics*[scale=0.28,angle=-90]{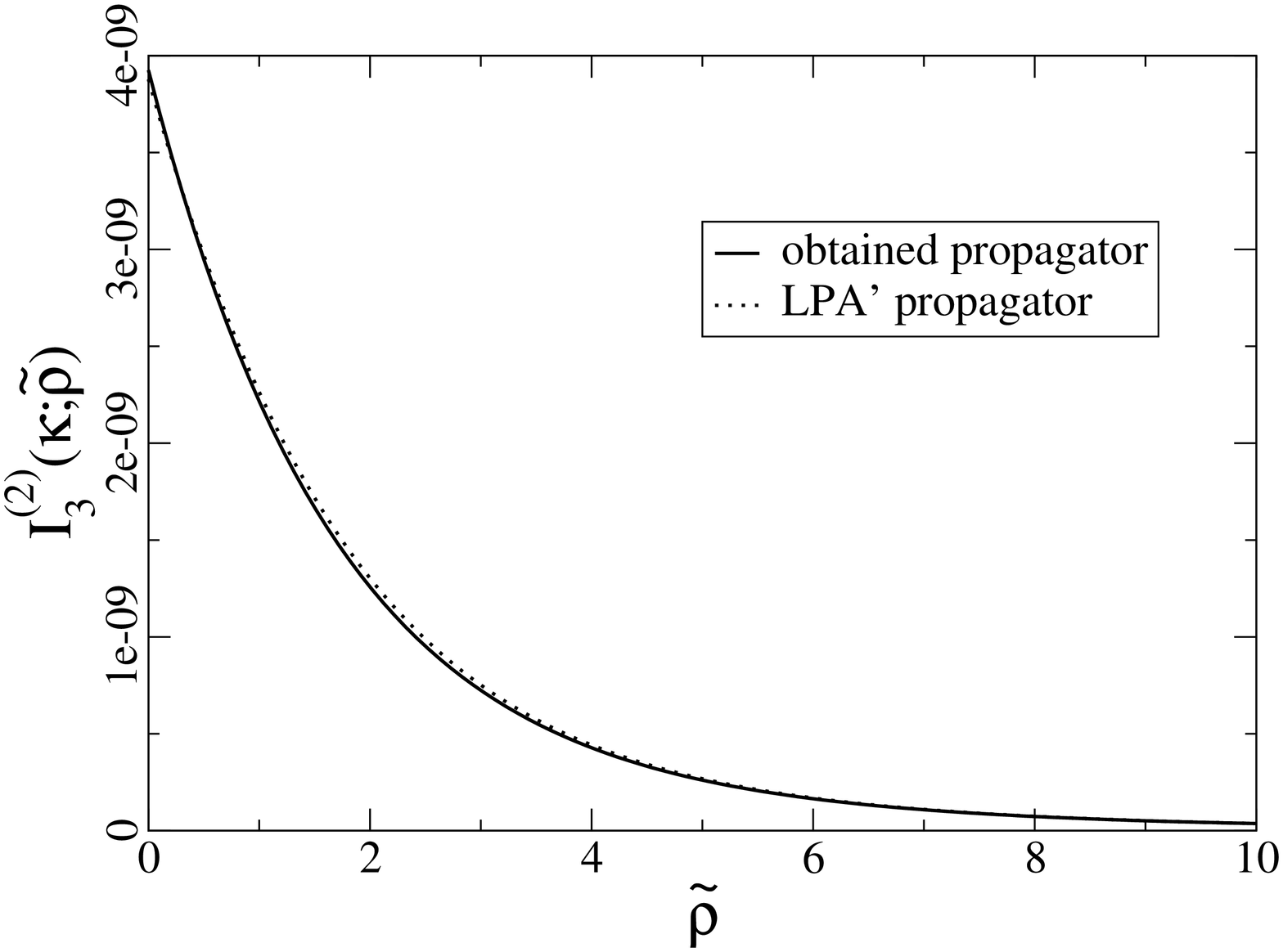}
\caption{ \label{I2_coc} Left: The ratio of the function
$I_{3}^{(2)}(\kappa;\tilde \rho)$ calculated with the  obtained numerical
propagator and with the approximate LPA' propagator (as explained in the
text), as a function of $\tilde \rho$, for different values of $\kappa/u$. Right: The function $I_{3}^{(2)}(\kappa;\tilde \rho)$ as a
function of $\tilde\rho$, for $\kappa/u=3\times 10^{-4}$, calculated with the
approximate propagator (dotted line) and with the obtained numerical
propagator (solid line). }
\end{figure}

 In
fig.~\ref{I2_coc} we plot the ratio of the function
$I^{(2)}_3(\kappa;\tilde\rho)$ ($\tilde\rho\sim\rho/\kappa$, see. eq.~(\ref{tilderho}))  calculated with the propagator obtained from the
numerical integration of the flow equation divided by the function  given by eq.~(\ref{In-anal}). One can see that the smaller
the value of $\kappa$, the larger the difference, and that the main
error is for values of $\tilde \rho$ around the minimum $\tilde\rho_{min}$ of the
effective potential ($\tilde\rho_{min}$ goes from 1.8 to  3 as $\kappa$ runs from $\Lambda$ to 0).  Nevertheless,   the difference stabilizes for
small enough $\kappa$ and it never exceeds  4\%. The right panel of
fig.~\ref{I2_coc}  shows the comparison of the two curves in the
worst situation, i.e., for
  small values of
$\kappa$, as a function of $\tilde \rho$: the curves are hardly distinguishable.

\begin{figure}[t]
\begin{center}
\includegraphics*[scale=0.28,angle=-90]{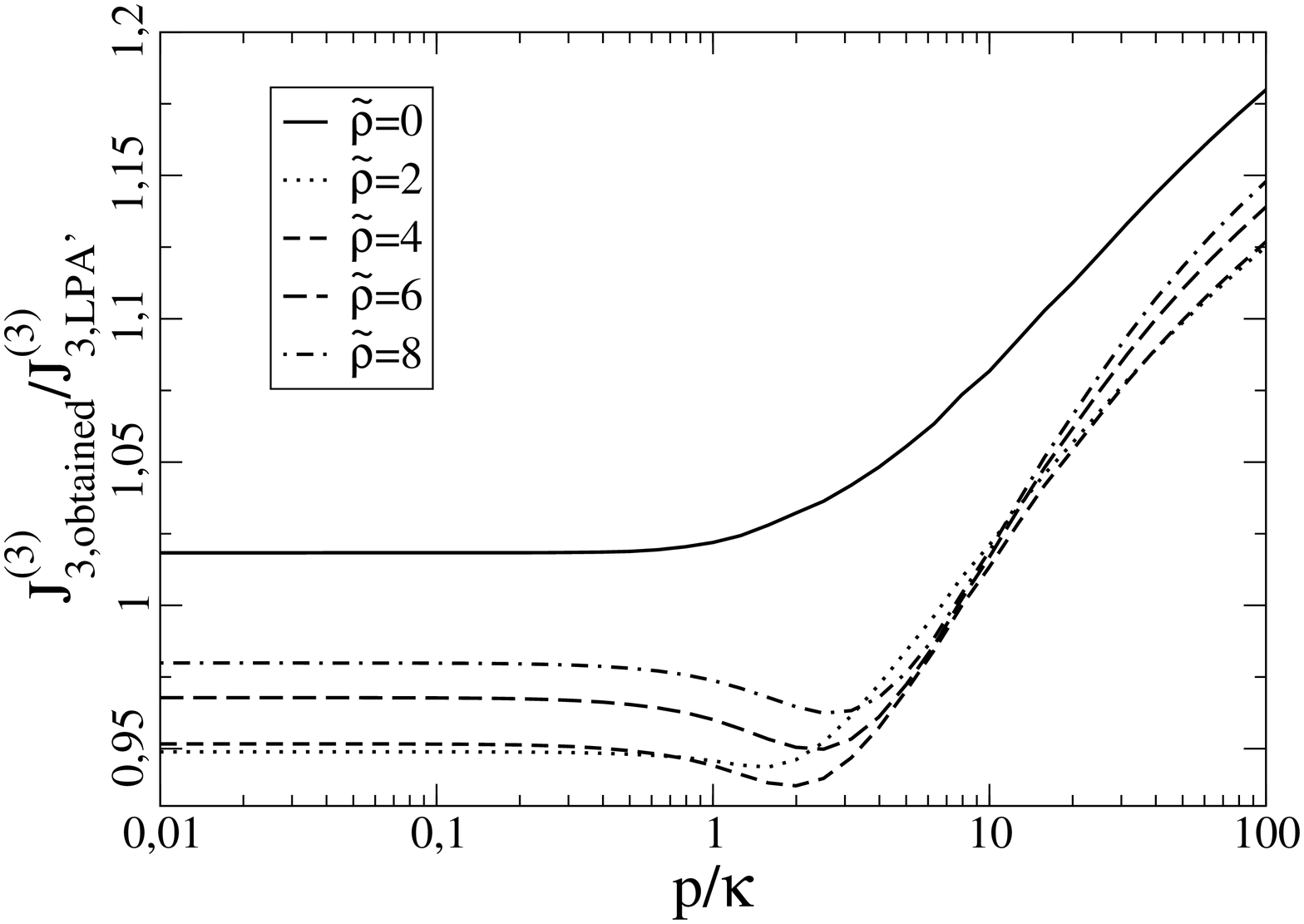}
\includegraphics*[scale=0.28,angle=-90]{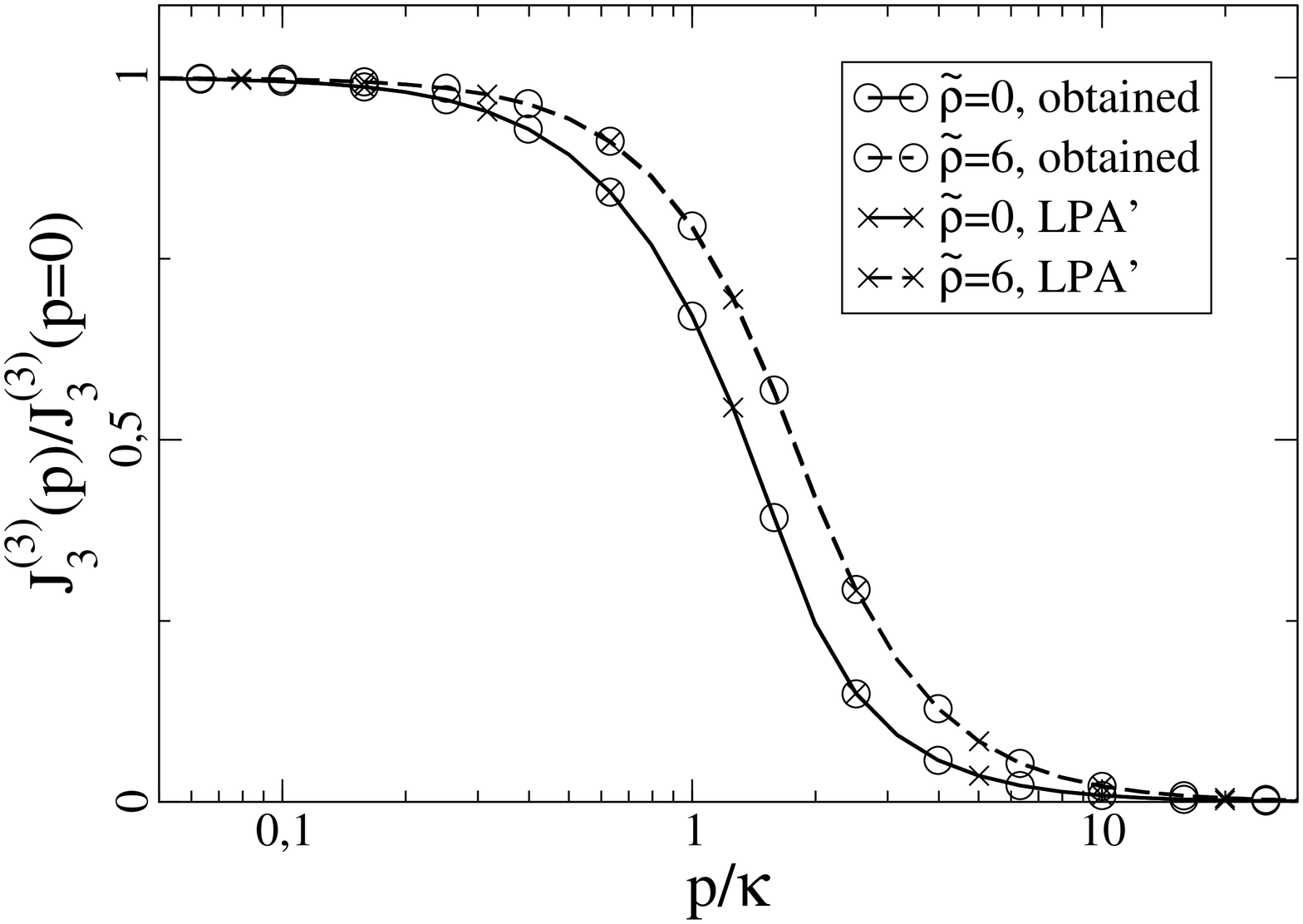}
\end{center}
\caption{ \label{J_comp} Left: The ratio of the function
$J^{(3)}_3(p;\kappa;\tilde\rho)$ calculated with the  obtained numerical
propagator and with the approximate LPA' propagators, as a function of
$p/\kappa$, for $\kappa/u=3\times 10^{-4}$ and for different
values of $\tilde \rho$.  Right: The  function $J^{(3)}_3(p;\kappa;\tilde\rho)/J^{(3)}_3(p=0;\kappa;\tilde\rho)$
calculated with the obtained numerical propagator compared to  that
calculated with the approximate propagators, as a function of
$p/\kappa$, for $\kappa/u=3\times 10^{-4}$ and for different values of
$\tilde\rho$. }
\end{figure}

The same analysis is repeated for  $J_3^{(3)}$. Again, it is  only for small values of $\kappa$ that the
two functions differs. In the left panel of fig.~\ref{J_comp} we display the ratio
of the function $J_3^{(3)}$ calculated respectively with the  obtained  (numerator) and the
approximate LPA' (denominator) propagators, for $\kappa/u = 3\times 10^{-4}$, for various
values of $\tilde\rho$. The difference can be large, but   only
 in the region ($p \gg \kappa$) where the function $J_3^{(3)}$ itself is very
small.  In the region
where the function is non negligeable, the difference between the two calculations never
 exceeds 5\%. As   was the case for  $I_3^{(2)}$, the largest error occurs for values of $\tilde\rho$ near the minumum of the potential. In the right panel of fig.~\ref{J_comp}, we plot
the two functions  for the same values of $\kappa$ and $\tilde \rho$ as in the left panel:  the
difference between the two calculations of $J_3^{(3)}$ is invisible on such a plot.

\section{Conclusions and perspectives}

We have demonstrated in this paper that the method proposed in
\cite{Blaizot:2005xy} allows for concrete numerical applications. We
have calculated the self-energy of the scalar model, at the LO of
the approximation scheme, at criticality, at zero external field, 
in $d=3$, and
have obtained accurate results over the whole range of momenta. 
Already at this level of approximation the results obtained compare 
well with those of more elaborate techniques. Worth emphasizing is 
the fact that the  scaling behavior of the self-energy is 
accurately reproduced: not only do we get a reasonable estimate of 
the anomalous dimension, but the entire dependence of the self-
energy on the momentum and the cut-off follows accurately the 
expected scaling behavior. 
To our knowledge, this is the first time that an approximate 
solution of the NPRG flow equations  is constructed with these 
properties. 

In the present paper, whose main objective was to confirm the applicability of the method to a concrete calculation, we solved approximately  the flow equation
(\ref{2pointclosed}). However, 
several tests suggest  that this approximate solution  differs  in fact very
little from the complete   solution of (\ref{2pointclosed}). 
Of course, a definite statement concerning the error made in the present calculation can only come
from a comparison with the exact solution. This, we believe, is within reach. Similarly, work is in progress to test the convergence of the procedure by calculating the next-to-leading order contribution. 

The method of ref.~\cite{Blaizot:2005xy}  builds on our previous works on the same subject \cite{Blaizot:2005wd,Blaizot:2006vr}. The results  presented in this paper indicate that it is both conceptually simpler, and numerically more accurate, than the method which we have developed in \cite{Blaizot:2005wd,Blaizot:2006vr}. It offers the possibility of 
 applications to a variety of  non-perturbative problems, where the knowledge of the momentum dependence of $n$-point functions is necessary. Even the approximate treatment presented in this paper could constitute an interesting starting point in situations where  only a semi-quantitative description would be valuable.

\acknowledgements
We would like to thank  Hugues Chat\'e, Bertrand Delamotte and 
Diego Guerra for many fruitful discussions. Ram\'on M\'endez-Galain 
and Nicol\'as Wschebor are grateful for the hospitality of  the 
ECT* in Trento  where part of this work has been carried out.

\appendix

\section{The $p=0$ sector}

\label{LPA}

As discussed in the main text, our approximate solution of eq.~(\ref{2pointclosed}) builds on the prior determination of quantities that are independent of momentum. These are calculated using a variant of the derivative expansion that we describe in this appendix. 
The  derivative expansion is usually \cite{Berges02}  formulated in terms of an ansatz for the running effective action $\Gamma_\kappa[\phi]$, including terms up to a given number of derivatives of the field.
Its leading order, the so-called local potential approximation (LPA), assumes that the effective action has the form:
\beq\label{gammaLPA}
\Gamma_\kappa^{LPA}[\phi]=\int d^dx \left\{
\frac{1}{2}\partial_{\mu}\phi_a\partial_{\mu}\phi_a+V_\kappa(\rho)\right\}.
\eeq
where the derivative term is simply the one appearing in the
classical action, and $V_\kappa(\rho)$ is the effective potential.
In the next-to-leading order
  (NLO), one assumes \cite{Wetterich93}
\beq \label{NLO}
\Gamma^{NLO}_\kappa[\phi]  =\int
d
^dx\left\lbrace
\frac{Z_\kappa (\rho)}{2}\partial_{\mu}\phi_a\partial_{\mu}\phi_a+V_\kappa(\rho)\right\rbrace .
\eeq
An interesting  improvement of the LPA, which we refer to as the
LPA',  is a simplified version of the NLO that
consists in ignoring the $\rho$-dependence of $Z_\kappa(\rho)$, i.e., in chosing $Z_\kappa = Z_\kappa(\rho_0)$ where $\rho_0$ is a given value of $\rho$, usually
taken to be the running
minimum of the potential. In the LPA' one solves simultaneously the flow equations for both
the effective potential $V_\kappa(\rho)$ (a partial differential
equation in $\kappa$ and $\rho$) and for $Z_\kappa$.
In this approximation, the inverse propagator
takes the form of eq.~(\ref{propGq}): $ G^{-1}_\kappa(q^2;\phi)={Z_\kappa
q^2+V_\kappa''(\phi)+R_\kappa(q)},$ with $V_\kappa''(\phi)=d^2V_\kappa/d\phi^2$. The LPA' allows for a non-trivial anomalous dimension, which is
determined from the cut-off dependence of $Z_\kappa$ (see eq.~(\ref{defZk}) and ref.~\cite{Wetterich93}).

The procedure followed in this paper to determine the field 
renormalisation constant $Z_\kappa$  differs slightly from that 
used in \cite{Blaizot:2005wd}. This is because, as explained in 
sect.~III, we need  the calculation of $Z_\kappa$ to be consistent 
with  the approximate eq.~(\ref{2pointclosed}) for the 2-point 
function. This is essential to get the proper scaling behavior of 
$\Gamma^{(2)}_\kappa(p;\rho)$ at small momenta. We set:
\begin{equation}\label{Zkapparho}
Z_\kappa(\rho)\equiv 1+\left.\frac{\partial \Sigma_{\kappa}(p;\rho)}{\partial p^2}\right|_{p=0},
\end{equation}
where $\Sigma_{\kappa}(p;\rho)$ is defined in  eq.~(\ref{defsigma}).
Notice that, using eq.~(\ref{def-Zk}), $ Z_\kappa(\rho_0)=Z_\kappa$.
The flow equation obeyed by $Z_\kappa(\rho)$ reads \begin{equation}\label{flowZrho}
\kappa \partial_\kappa Z_\kappa(\rho)=\left.\frac{\partial J_d^{(3)}(p^2,\rho)}{\partial p^2}\right|_{p=0}
\left(\frac{\partial^3 V}{\partial \phi^3} \right)^2+2 I_d^{(3)}(\rho)\frac{\partial^3 V}{\partial \phi^3}
\frac{\partial Z_\kappa(\rho)}{\partial \phi}-\frac{1}{2} I_d^{(2)}(\rho)\frac{\partial^2 Z_\kappa(\rho)}{\partial \phi^2},
\end{equation}
which follows immediately from eq.~(\ref{2pointclosed}) for $\Sigma_\kappa(p;\rho)$. Knowing the solution of this equation we can calculate $\eta_\kappa$ from eqs.~(\ref{defZk}) and (\ref{def-Zk}). At this point, it is convenient to choose $\rho_0=0$. Then the expression of $\eta_\kappa$ that one deduces from eq.~(\ref{flowZrho}) simplifies into:
\begin{equation}\label{eqneta}
\eta_\kappa=\frac{1}{2} I_d^{(2)}(\rho=0)\frac{1}{Z_\kappa}\left.\frac{\partial^2 Z_\kappa(\rho)}{\partial \phi^2}\right|_{\rho=0}.
\end{equation}
Since $I_d^{(2)}(\rho=0)$ depends explicitly on $Z_\kappa$ and
$\eta_\kappa$ (see eq.~(\ref{In-anal})), eq.~(\ref{eqneta}) is in
fact a self-consistent equation for $\eta_\kappa$. The fact that a 
derivative of $Z_\kappa(\rho)$ enters eq.~(\ref{eqneta}) demands 
the simultaneous resolution of eq.~(\ref{flowZrho}) for small 
finite values of $\rho$.

The  solution of the LPA' is well
documented in the literature (see e.g. \cite{Berges02,Canet02}).
In practice, we work with  dimensionless quantities. We set:
\begin{eqnarray} \label{adim}
v_\kappa (\tilde \rho)\equiv K_d^{-1} \kappa^{-d} V_\kappa (\rho),\qquad
\chi(\tilde \rho)\equiv  \frac{Z_\kappa(\rho)}{Z_\kappa},
\end{eqnarray}
with \beq \label{tilderho}\tilde \rho \equiv  K_d^{-1} \; Z_\kappa \; \kappa^{2-d}
\;\rho, \eeq and $K_d$ is given after eq.~(\ref{defm2k}). We solve
the equation for the derivative of the potential
 with respect to $\tilde\rho$, i.e., $w_\kappa(\tilde\rho)\equiv \partial_{\tilde\rho}
v_\kappa(\tilde\rho)$, rather than that for the effective potential itself. This reads (from now on we stick to $d=3$):
\beq\label{flow_of_w}
\kappa\partial_\kappa w_\kappa\!=\! -(2\!-\!\eta_\kappa) w_\kappa + (1+\!\eta_\kappa) \tilde\rho w'_\kappa
 -\! \left( 1\!-\!\frac{\eta_\kappa}{5} \right) \!\left( \frac{(N\!-\!1) w'_\kappa}{(1+w_\kappa)^2} \!+\!
 \frac{3w'_\kappa + 2\tilde\rho w''_\kappa}{(1+w_\kappa+2 \tilde\rho w'_\kappa)^2} \right),\nonumber\\
\eeq
where $w'_\kappa=\partial_{\tilde\rho} w_\kappa(\tilde\rho)$, $w''_\kappa=\partial_{\tilde\rho}^2 w_\kappa(\tilde\rho)$. Eq.~(\ref{flow_of_w}) is solved starting from the initial condition at
$\kappa=\Lambda$:
\beq\label{wLambdadez}
w_\kappa(\tilde\rho,\kappa=\Lambda) = \hat m_\Lambda^2 + \hat g_\Lambda \tilde\rho,
\eeq
where $ \hat m_\Lambda$ and $ \hat g_\Lambda$ are related to the parameters $r$ and $u$ of the classical action (\ref{eactON}) by
\beq\label{relclas}
 \hat m_\Lambda^2 =\frac{r}{\Lambda^2},\qquad \hat g_\Lambda = \frac{u }{\Lambda}\frac{K_3}{3},
\eeq
and the parameter $r$ is adjusted to be at criticality.
Together with eq.~(\ref{flow_of_w}), we solve the equation for $\chi_\kappa(\tilde\rho>0)$, which  reads
\beq
\kappa\del_\kappa \chi_\kappa&=&\eta_\kappa \chi_\kappa+(1+\eta_\kappa)\tilde\rho \chi_\kappa'-2\tilde\rho \frac{(3w'_\kappa+2\tilde\rho w''_\kappa)^2}{(1+w'_\kappa+\tilde\rho w''_\kappa)^4}
\nonumber\\ &+&8\tilde\rho\chi_\kappa' (1-\frac{\eta_\kappa}{5})\frac{(3w'_\kappa+2\tilde\rho w''_\kappa)}{(1+w'_\kappa+2\tilde\rho w''_\kappa)^3}-(1-\frac{\eta_\kappa}{5})\frac{\chi_\kappa'+2\tilde\rho\chi_\kappa''}{(1+w'_\kappa+2\tilde\rho w''_\kappa)^2},
\eeq
where $\chi_\kappa'=d\chi_\kappa/d\rho$. The initial condition is $\chi_\kappa(\tilde\rho=0)=1$ for all $\kappa$, which follows from the definition of $Z_\kappa$, eq.~(\ref{def-Zk}).
Finally,  for $\eta_\kappa$ we have simply:
\beq
\eta_\kappa=\frac{\chi'_\kappa(0)}{(1+w_\kappa'(0))^2+\chi_\kappa'(0)/5}.
\eeq

\begin{figure}[t!]
\begin{center}
\includegraphics*[scale=0.28,angle=-90]{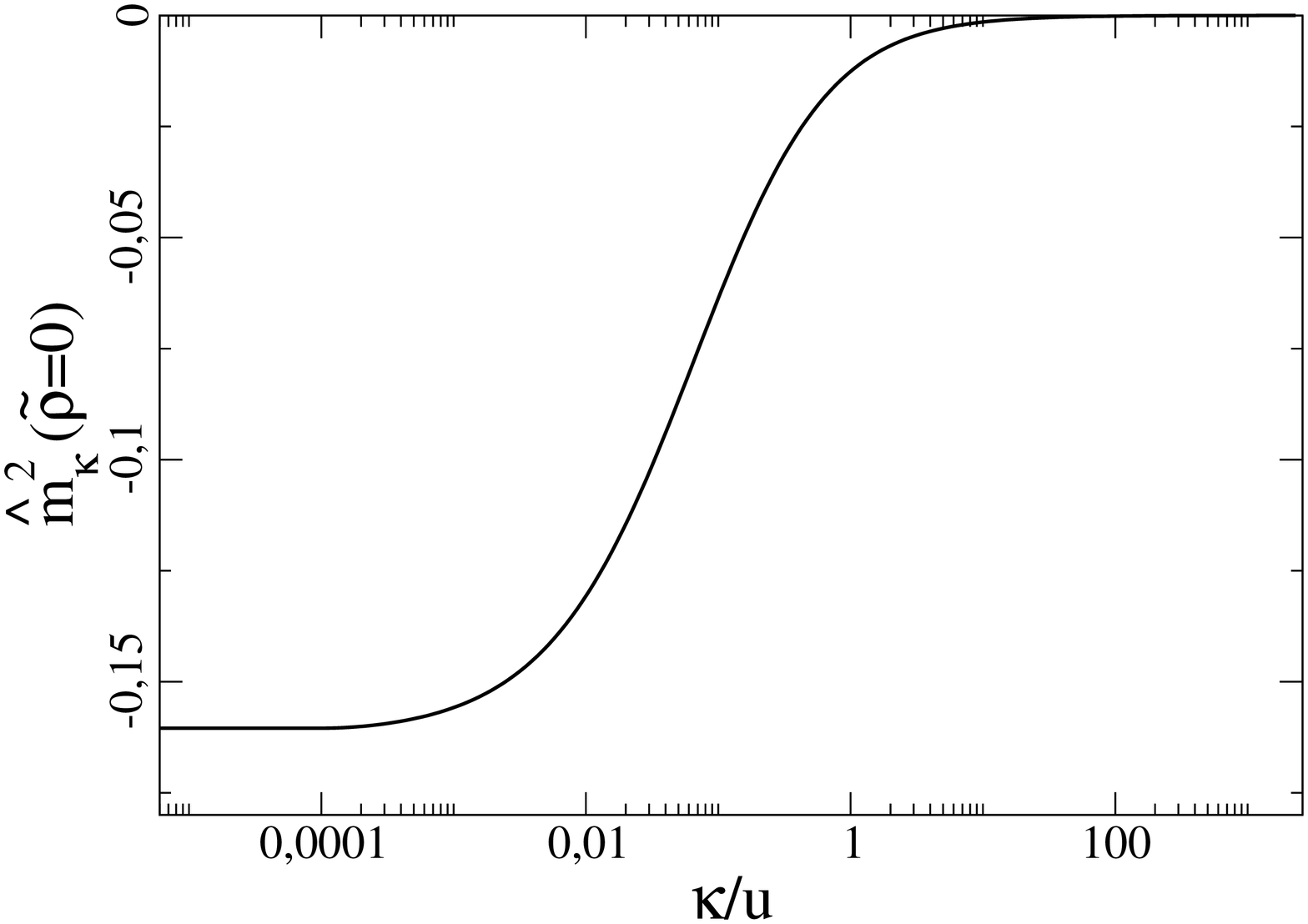}
\includegraphics*[scale=0.28,angle=-90]{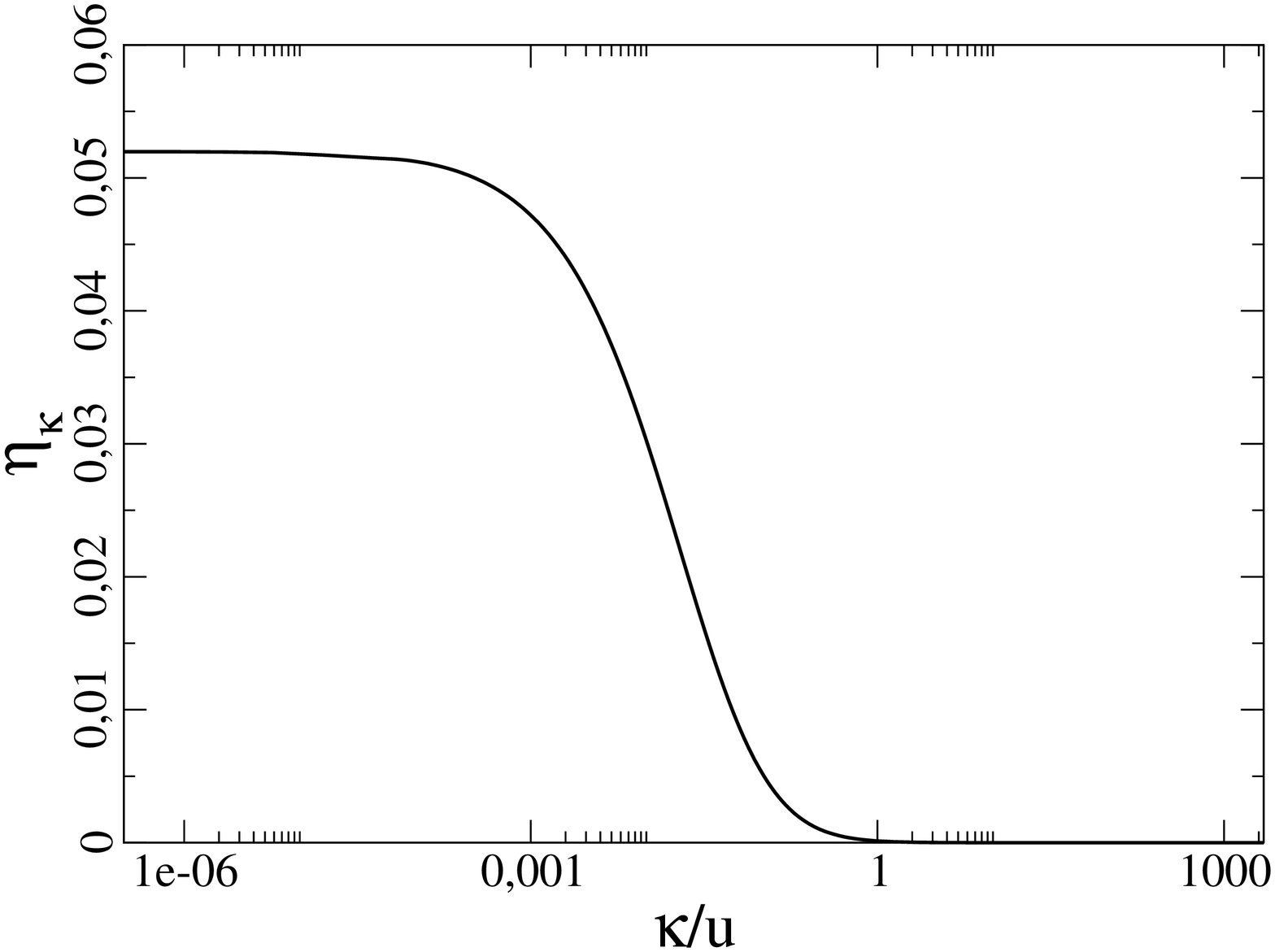}
\end{center}
\caption{\label{m2k} The dimensionless mass $\hat m^2_\kappa(\tilde\rho=0)=\partial_{\tilde\rho} v_\kappa(\tilde\rho=0)$ (left) and the anomalous dimension $\eta_\kappa$ (right) as a function of
$\kappa/u$. These quantities were  obtained by solving the LPA' equations, with $u/\Lambda=3.54\times 10^{-4}$ and the parameter $r$  adjusted to be at criticality. }
\end{figure}

\begin{figure}[t!]
\begin{center}
\includegraphics*[scale=0.40,angle=-90]{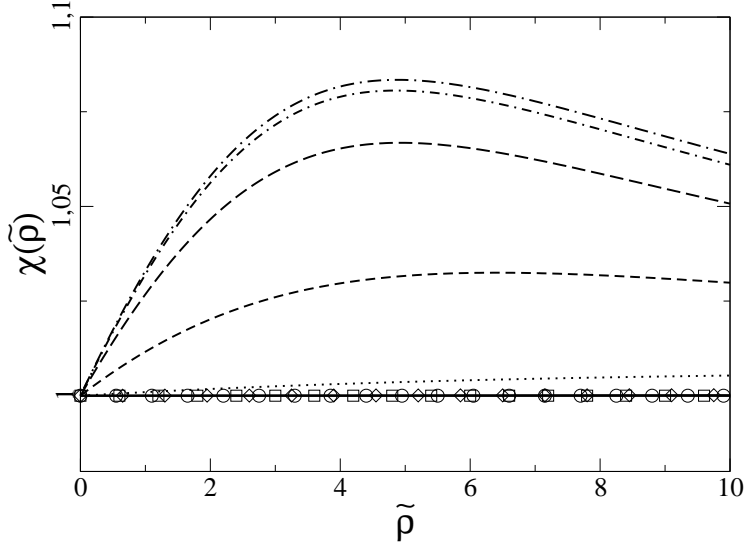}
\end{center}
\caption{\label{chi} The dimensionless function $\chi(\tilde\rho)$ defined in eq.~(\ref{adim}), for various values of $\kappa/u=3\times 10^{n}$, with $n=2$ (circles), $n=1$ (squares), $n=0$ (diamonds), $n=-1,\cdots,-5$ from bottom to top. }
\end{figure}

For the sake of illustration, we  present in fig.~\ref{m2k} the LPA' solutions for  $\hat m^2_\kappa(\tilde\rho=0)$
(defined in eq.~(\ref{defm2k})) and $\eta_\kappa$, as a function of $\kappa/u$. The calculations have been done with $u/\Lambda=3.54\times 10^{-4}$,
but the curves  are independent of this choice, provided $u/\Lambda$ remains small. One can verify that the  the crossover between the UV and IR regimes occurs around $\kappa_c \sim u/10$. The fixed point value of $\eta_\kappa$ is
 $\eta^*=\eta_{\kappa \to 0} \approx 0.05220$. Fig.~\ref{chi} illustrates the $\rho$-dependence of the renormalization factor $Z_\kappa(\rho)$ (see eq.~(\ref{adim})). This dependence is completely negligible when $\kappa\simge \kappa_c\sim u/10$, and never exceeds $8\%$.
 
\section{The functions $I_3^{(2)}(\kappa;\rho)$ and $J_3^{(3)}(p;\kappa;\rho)$}
\label{functions}

In this appendix we provide  details about the functions
$I_{3}^{(2)}(\kappa;\rho)$ and $J_{3}^{(3)}(p;\kappa;\rho)$
calculated with the propagators (\ref{propGq}) and (\ref{propGpq})
respectively.

\begin{figure}[t]
\begin{center}
\includegraphics*[scale=0.40,angle=-90]{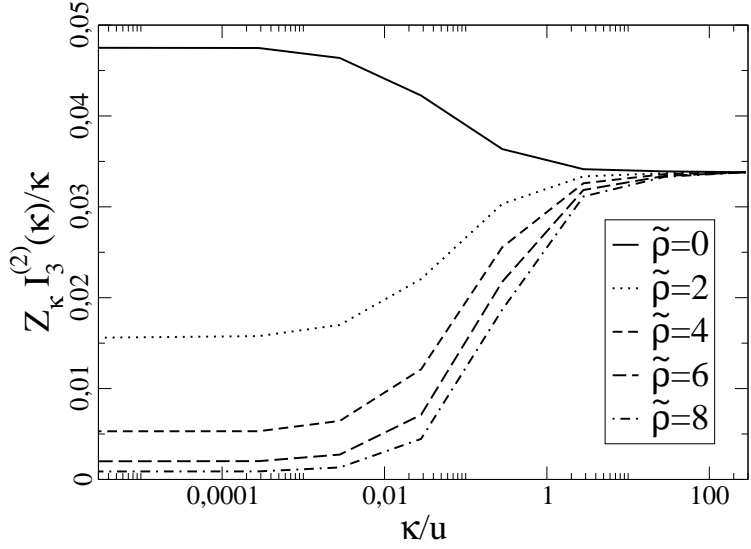}
\end{center}
\caption{ \label{I2sk_5z} The function $Z_\kappa
I_{3}^{2}(\kappa;\tilde \rho)/\kappa$ as a function of $\kappa/u$,
for
 different values of $\tilde \rho$.}
\end{figure}

Consider first the function $I_{3}^{(2)}(\kappa;\rho)$, defined in
eq.~(\ref{defI}), and whose explicit expression is given in
eq.~(\ref{In-anal}). The variation of $I_{3}^{(2)}(\kappa;\rho)$
with $\kappa$ is  dominated by the explicit linear $\kappa$
dependence and the $\kappa$-dependence of the renormalization factor
$Z_\kappa$. The function \beq \frac{Z_\kappa
I_3^{(2)}(\kappa;\rho)}{\kappa} = 2 K_3 \frac{1}{(1+\hat
m^2_\kappa(\tilde \rho))^2} \left(1-\frac{\eta_\kappa}{5}\right),
\eeq  displayed in fig.~\ref{I2sk_5z}, illustrates the remaining
dependence on $\kappa$ and $\tilde\rho$.

Consider next  $J_3^{(3)}(p;\kappa;\rho)$, defined in
eq.~(\ref{defJ}). Using the LPA' propagators of eqs.~(\ref{propGq})
and (\ref{propGpq}) one can calculate it analytically. One first
makes the changes of variables $\bar p=p/\kappa, \bar q = q/\kappa$
and $\cos \gamma=p.q/p\;q$, and then perform the integral over the
remaining angular variables. One gets then \begin{eqnarray}
J_3^{(3)}(p;\kappa;\rho)
&=&
\frac{1}{\kappa Z_\kappa^2(2\pi)^2}\frac{1}{(1+\hat m_\kappa^2)^2}
\int_0^1 d\bar q \bar q^{2}\int_{-1}^{1}d(\cos \gamma)
 \nonumber \\
\hspace{-.8cm}
 \times\, &&
\hspace{-.8cm}
\frac{(2-\eta+\eta \bar q^2)}{\Theta(1\!-\!\bar q^2\!-\!\bar p^2\!+\!2 \bar q \bar p \cos \gamma)\!+\!
(\bar q^2\!+\!\bar p^2-2 \bar q \bar p \cos \gamma)
\Theta(\bar q^2\!+\!\bar p^2\!-\!2 \bar q \bar p \cos \gamma\!-\!1)\!+\!\hat m_\kappa^2}.  \nonumber \\
\end{eqnarray}
To perform the integral over $\cos \gamma$ one needs to consider the
various domains defined by  the $\Theta$ functions. It is then
convenient to separate the calculation in two different regions:
$\bar p>2$ and $\bar p\leq 2$,  and this for the two possible signs
of $\hat m_\kappa^2$ (see also \cite{Blaizot:2005wd}). The
calculation is then done by first  integrating over $\cos \gamma$; the
remaining integration over $\bar q$ can be done by making first an
integration by parts to get a rational function, that is then
decomposed into simple fractions. One finally gets (the dependence
on $\tilde\rho$ is entirely contained in $\hat m_\kappa
(\tilde\rho)$ and is not written out explicitly):

a) $\bar p>2, \hat m_\kappa^2<0 $.
\begin{eqnarray}\label{Jlarge}
 &&\hspace{-.2cm}J_3^{(3)}(p;\kappa;\tilde\rho)
=\frac{1}{\kappa Z_\kappa^2(2\pi)^2}\frac{1}{(1+\hat m_\kappa^2)^2}\left\lbrace 2
+\frac{\eta}{2}\left(-\frac{5}{3}+\bar p^2-3\hat m_\kappa^2\right) \right.
\nonumber \\
&&+\frac{1}{2\bar p}\left[-1+\frac{\eta}{4}+\left(\bar p+\sqrt{-\hat m_\kappa^2}\right)^2
\left(1-\frac{\eta}{2}
+\frac{\eta}{4}\left(\bar p+\sqrt{-\hat m_\kappa^2}\right)^2\right)\right]
\log\left(\frac{\bar p-1+\sqrt{-\hat m_\kappa^2}}{\bar p+1+\sqrt{-\hat m_\kappa^2}}
\right) \nonumber \\
&&+\left.\frac{1}{2\bar p}\left[-1+\frac{\eta}{4}+\left(\bar p-\sqrt{-\hat m_\kappa^2}
\right)^2\left(1-\frac{\eta}{2}
+\frac{\eta}{4}\left(\bar p-\sqrt{-\hat m_\kappa^2}\right)^2\right)\right]
\log\left(\frac{\bar p-1-\sqrt{-\hat m_\kappa^2}}{\bar p+1-\sqrt{-\hat m_\kappa^2}}
\right) \right\rbrace \nonumber \\
&&=\frac{1}{\kappa Z_\kappa^2(2\pi)^2}\frac{2}{(1+\hat m_\kappa^2)^2}\left\lbrace
\frac{4}{\bar p^2}\left(\frac{1}{3}-\frac{\eta}{15}\right)+\frac{4}{\bar p^4}\left(
\frac{1}{15}-\frac{\eta}{105}
-\frac{\hat m_\kappa^2}{3}+\frac{\eta\hat m_\kappa^2}{15}\right)+
\mathcal{O}(1/(\bar p^6)) \right\rbrace . \nonumber \\
\end{eqnarray}
b) $\bar p\le 2, \hat m_\kappa^2<0$.
\begin{eqnarray}\label{Jsmall}
J_3^{(3)}(p;\kappa;\tilde\rho)
&=&
\frac{1}{\kappa Z_\kappa^2(2\pi)^2}\frac{1}{(1+\hat m_\kappa^2)^2}\left\lbrace
-1+\frac{\eta}{4}+\frac{\eta\hat m
_\kappa^2}{4}+\bar p\left(\frac{3}{2}
-\frac{\eta}{8}-\frac{7\eta\hat m_\kappa^2}{8}\right)
-\frac{3\eta}{4}\bar p^2\right. \nonumber \\
&+&\frac{25\eta}{48}\bar p^3+\frac{1}{1+\hat m_\kappa^2}\left(\frac{4}{3}
-\frac{4\eta}{15}-\bar p+\frac{\eta}{3}\bar p^2
+\left(\frac{1}{12}-\frac{\eta}{6}\right)\bar p^3+\frac{\eta}{120}\bar p^5\right)
\nonumber \\
&+&\hspace{-.2cm}
\frac{1}{2\bar p}\left[1-\frac{\eta}{4}
-\left(\bar p+\sqrt{-\hat m_\kappa^2}\right)^2
\left(1-\frac{\eta}{2}+\frac{\eta}{4}
\left(\bar p+\sqrt{-\hat m_\kappa^2}\right)^2\right)\right]
\log \left(\frac{\bar p+1+\sqrt{-\hat m_\kappa^2}}{1+\sqrt{-\hat m_\kappa^2}}\right)
\nonumber \\
&+&\left.\hspace{-.2cm}\frac{1}{2\bar p}\left[1-\frac{\eta}{4}-
\left(\bar p-\sqrt{-\hat m_\kappa^2}\right)^2\left(1-\frac{\eta}{2}+\frac{\eta}{4}
\left(\bar p-\sqrt{-\hat m_\kappa^2}\right)^2\right)\right]
\log\left(\frac{\bar p+1-\sqrt{-\hat m_\kappa^2}}{1-\sqrt{-\hat m_\kappa^2}}\right)
\right\rbrace\nonumber \\
&=&
\frac{1}{\kappa Z_\kappa^2(2\pi)^2}\frac{1}{(1+\hat m_\kappa^2)^2}\left\lbrace
\frac{4}{3(1+\hat m_\kappa^2)}\left(1-\frac{\eta}{5}\right)-\frac{2}{3(1+\hat m_\kappa^2)^2}\bar p^2 \right.\nonumber \\
&+&\left. \frac{2+\eta-2\hat m_\kappa^2+\eta \hat m_\kappa^2}{6(1+\hat m_\kappa^2)^3}\bar p^3
-\frac{2(1+\eta-5\hat m_\kappa^2+\eta \hat m_\kappa^2)}{15(1+\hat m_\kappa^2)^4}\bar p^4+\mathcal{O}(\bar p^5)\right\rbrace.
\end{eqnarray}

c) $\bar p>2, m_\kappa^2\ge 0$.
\begin{eqnarray}\label{Jlarge2}
&&\hspace{-.2cm}J_3^{(3)}(p;\kappa;\tilde\rho) =\frac{1}{\kappa
Z_\kappa^2(2\pi)^2}\frac{1}{(1+\hat m_\kappa^2)^2}\left\lbrace 2
+\frac{\eta}{2}\left(-\frac{5}{3}+\bar p^2-3\hat m_\kappa^2\right) \right. \nonumber \\
&&+\frac{1}{\bar p}\left[\left(-1+\frac{\eta}{4}+(\bar p^2-\hat m_\kappa^2)\left(1-\frac{\eta}{2}
+\frac{\eta}{4}(\bar p^2-\hat m_\kappa^2)\right)-\eta\hat m_\kappa^2\bar p^2\right)\frac{1}{2}\log\left(\frac{(\bar p-1)^2+\hat m_\kappa^2}{(\bar p+1)^2+\hat m_\kappa^2}
\right) \right.\nonumber\\
&&\left.\left.-2\hat m_\kappa \bar p \left(1-\frac{\eta}{2}+\frac{\eta}{2}(\bar p^2-\hat m_\kappa^2)\right)\left(
\mathrm{Arctan} \left(\frac{\hat m_\kappa}{\bar p-1}\right)-\mathrm{Arctan} \left(\frac{\hat m_\kappa}{\bar p+1}\right)\right)\right]\right\}\nonumber \\
&&=\frac{1}{\kappa Z_\kappa^2(2\pi)^2}\frac{1}{(1+\hat
m_\kappa^2)^2}\left\lbrace
\frac{4}{\bar p^2}\left(\frac{1}{3}-\frac{\eta}{15}\right)\right. \nonumber\\
&&\left.+\frac{1}{105\,\bar p^4}\left(7-35\hat m_\kappa^2+\eta(-1+7\hat m_\kappa^2)\right)+\mathcal{O}(1/(\bar p^6)) \right\rbrace \nonumber \\
\end{eqnarray}

d) $\bar p\leq 2, m_\kappa^2\ge 0$.
\begin{eqnarray}\label{largeJ4}
J_3^{(3)}(p;\kappa;\tilde\rho) &=&\frac{1}{\kappa
Z_\kappa^2(2\pi)^2(1+\hat m_\kappa^2)^2}\left\lbrace
-1+\frac{\eta}{4}+\frac{\eta \hat m_\kappa^2}{4}+\bar
p\left(\frac{3}{2}-\frac{\eta}{8}-\frac{7\eta \hat
m_\kappa^2}{8}\right)
-\frac{3\eta \bar p^2}{4} \right. \nonumber \\
&+& \frac{25\eta \bar p^3}{48}+\frac{1}{1+\hat m_\kappa^2}\left(\frac{4}{3}-\frac{4\eta}{15}-\bar p+\frac{\eta \bar p^2}{3}
+\frac{\bar p^3}{12}-\frac{\eta \bar p^3}{6}+\frac{\eta \bar p^5}{120} \right) \nonumber \\
&+&\frac{1}{\bar p}\left[\left(1-\frac{\eta}{4}-(\bar p^2-\hat m_\kappa^2)\left(1-\frac{\eta}{2}+\frac{\eta}{4}
(\bar p^2-\hat m_\kappa^2)\right)+\eta \hat m_\kappa^2 \bar p^2\right)\frac{1}{2}\log \left(\frac{(\bar p+1)^2+\hat m_\kappa^2}{1+\hat m_\kappa^2}\right) \right.\nonumber \\
&+&\left.\left.2\hat m_\kappa \bar p \left(1-\frac{\eta}{2}+\frac{\eta}{2}(\bar p^2-\hat m_\kappa^2)\right)
\left(\mathrm{Arctan} \left(\frac{\hat m_\kappa}{\bar p+1}\right)-\mathrm{Arctan} \left(\hat m_\kappa\right)\right)\right]\right\} \nonumber \\
&=&\frac{1}{\kappa Z_\kappa^2(2\pi)^2(1+\hat
m_\kappa^2)^2}\left\lbrace
\frac{4}{3(1+\hat m_\kappa^2)}\left(1-\frac{\eta}{5}\right)-\frac{2}{3(1+\hat m_\kappa^2)^2}\bar p^2 \right.\nonumber \\
&+&\left. \frac{2+\eta-2\hat m_\kappa^2+\eta \hat m_\kappa^2}{6(1+\hat m_\kappa^2)^3}\bar p^3
-\frac{2(1+\eta-5\hat m_\kappa^2+\eta \hat m_\kappa^2)}{15(1+\hat m_\kappa^2)^4}\bar p^4+\mathcal{O}(\bar p^5)\right\rbrace .
\end{eqnarray}

\begin{figure}[t]
\begin{center}
\includegraphics*[scale=0.28,angle=-90]{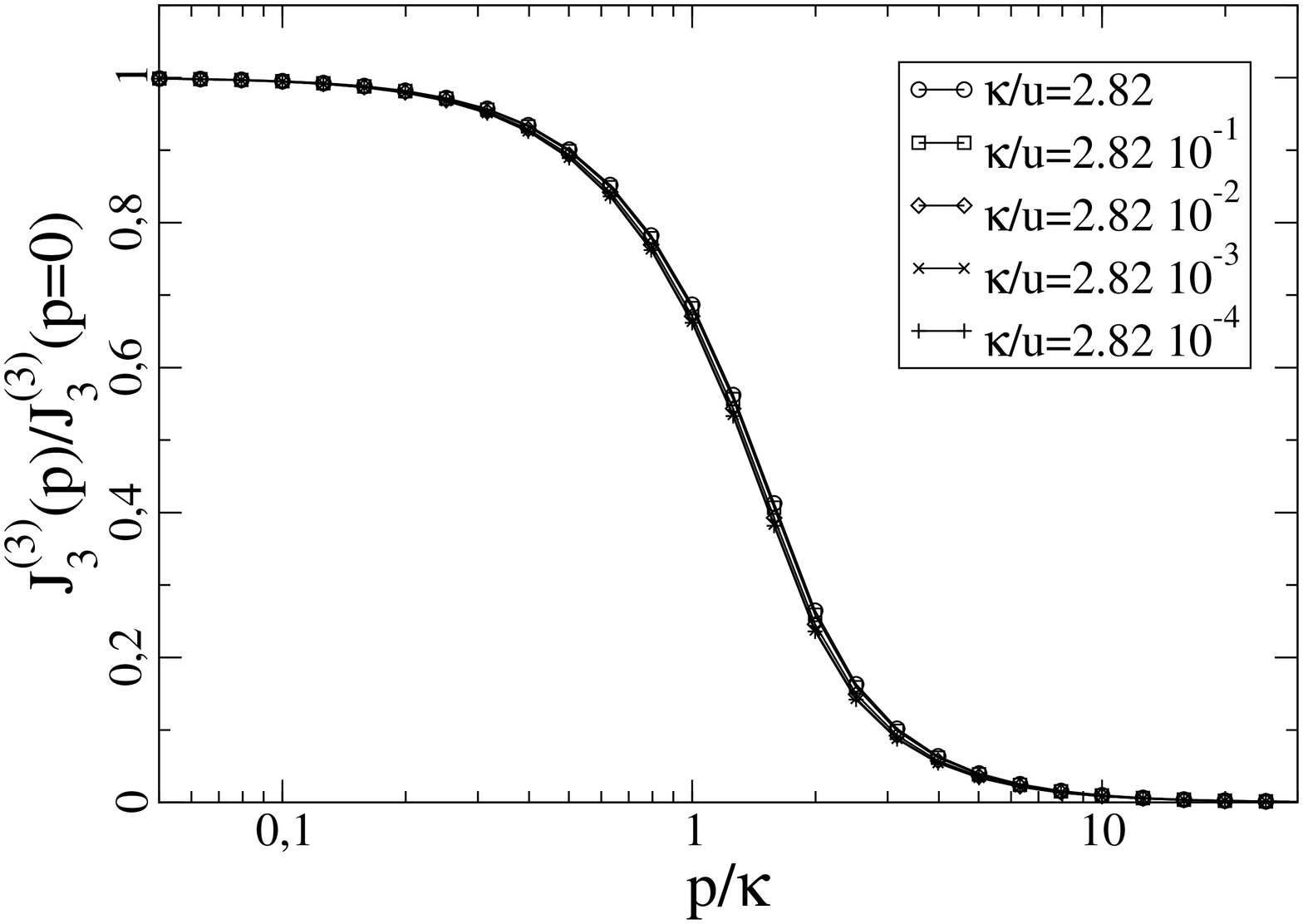}
\includegraphics*[scale=0.28,angle=-90]{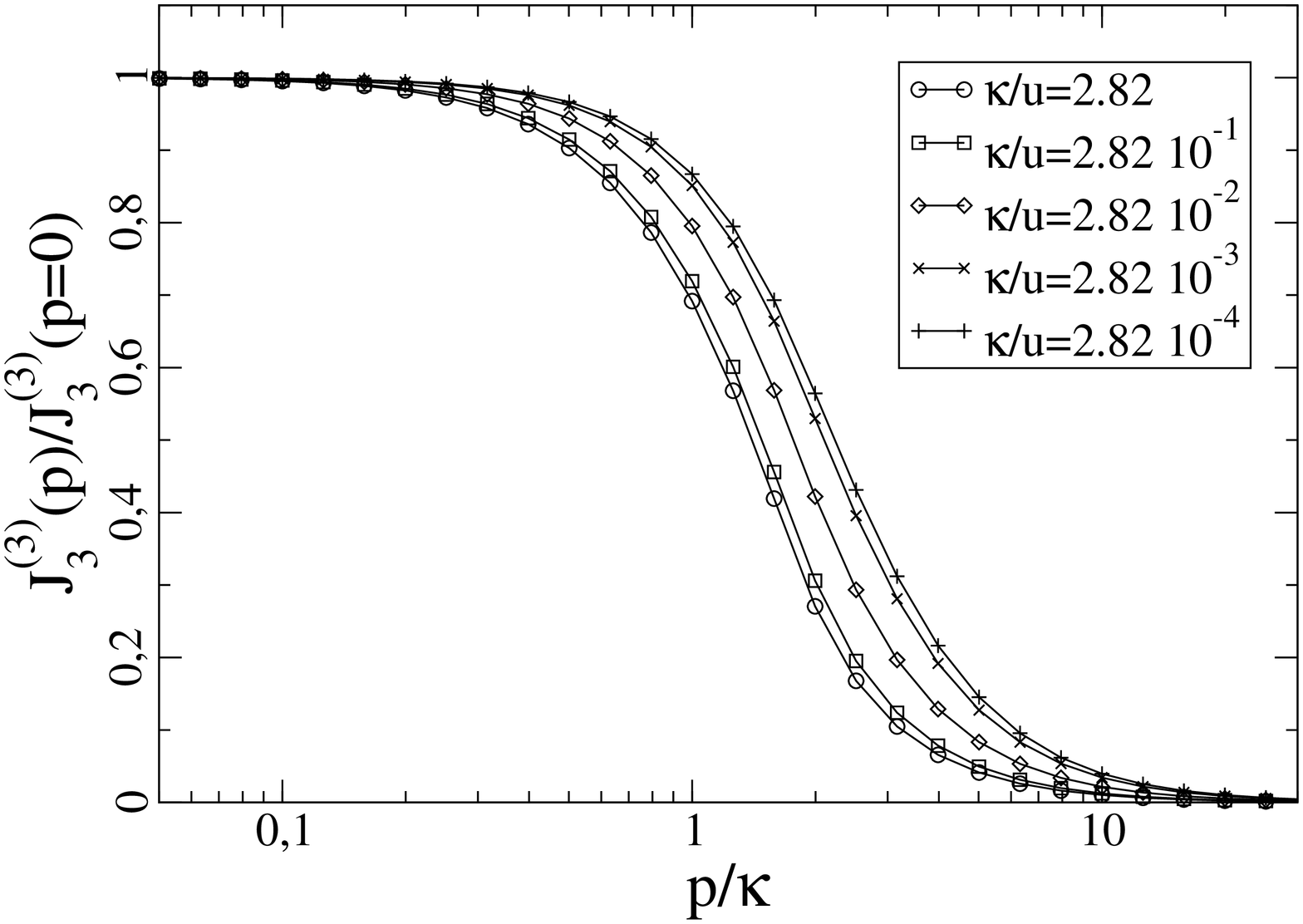}
\end{center}
\caption{ \label{an-spre-z0} The function $J_3^{(3)}(\kappa;p) /I_3^{(3)}(\kappa)$ for $\tilde \rho=0$ (left) and $\tilde\rho=6$ (right), as a function of  $(p/\kappa)$, for different values of $\kappa/u$. }
\end{figure}

The function $J_3^{(3)}(p;\kappa;\rho)/J_3^{(3)}(p=0;\kappa;\rho)$
is displayed in fig.~\ref{an-spre-z0} for the two values
$\tilde\rho=0$, and $\tilde\rho=6$. One sees that in both cases the
$p$-dependence is concentrated in the region $p\sim\kappa$:
$J_3^{(3)}(p;\kappa;\rho)$ is independent of $p$ when $p\simle
\kappa$, and it vanishes when $p\simge\kappa$, a property that has
also been exploited in \cite{Blaizot:2005wd} and  
\cite{Blaizot:2006vr}.
For $\tilde\rho=0$, $J_3^{(3)}(\kappa;p)/J_3^{(3)}(\kappa;p=0)$ is
essentially a function of $p/\kappa$ only. For $\tilde\rho=6$ some
residual dependence on $\kappa$ remains.

\section{Ultraviolet behavior of the self-energy}
\label{perturbative}

In this appendix we study the behavior of the self-energy
$\Sigma(p)$ for  $p\gg u$.  We show that the solution of
eq.~(\ref{2pointclosed}) reproduces the result of 2-loop perturbation theory, namely
$\Sigma(p)=u^2/(96\pi^2) \log(p/u)$, albeit with a
coefficient  in front of the logarithmic that differs by 8\%.

Consider first the exact flow equation for the 2-point function,
eq.~(\ref{gamma2champnonnul}), in vanishing external field (in this
appendix $\rho=0$ throughout). At order 0-loop (indicated by the
superscript $[0]$), this is simply:
\begin{equation}
\partial_\kappa \Gamma_{\kappa}^{(2)[0]}(p) = 0.
\end{equation}
This equation  has
the solution
\begin{equation}\label{prop0}
\Gamma_{\kappa}^{(2)[0]}(p)= p^2 ,
\end{equation}
where we used the initial condition at $\kappa=\Lambda$ that one deduces from  eq.~(\ref{eactON}), and adjusted the bare mass
 $r$ to be at criticality ($\Sigma_{\kappa=0}(p=0;\rho=0)=0$, yielding $r^{[0]}=0$).

To go to   1-loop, one uses, in the r.h.s. of
eq.~(\ref{gamma2champnonnul}), the 0-loop expressions for both the
propagator, $G_0(\kappa;p) = {1}/({p^2+R_\kappa(p)})$,  and  the
4-point function $\Gamma_{\kappa}^{(4)[0]}(p_1,p_2,p_3.p_4)=u$. One
gets\beq
\partial_\kappa \Gamma_{\kappa}^{(2)[1]}(p)&=&-\frac{u}{2}\;\int \frac{d^dq}{(2\pi)^d}
\frac{\partial_\kappa R_\kappa(q)}{(q^2+R_\kappa(q))^2}=\frac{u}{2}\partial_\kappa \int \frac{d^dq}{(2\pi)^d}
\frac{1}{q^2+R_\kappa(q)} \nonumber\\
\eeq
The integration is immediate; by imposing criticality and the initial condition at $\kappa=\Lambda$, one obtains
\begin{equation}\label{prop1}
\Gamma_{\kappa}^{(2)[1]}(p)=p^2+\frac{u}{2}\; \int
\frac{d^dq}{(2\pi)^d}\left\{\frac{1}{q^2+R_\kappa(q)}-\frac{1}{q^2}\right\},
\end{equation}
giving a self-energy which is in fact independent of the 
momentum $p$.

 The 1-loop expression for  the 4-point function, which will be needed shortly, is obtained similarly:
 \begin{eqnarray}\label{4point1a}
&&\!\!\!\!\!\!\!{\partial_\kappa
\Gamma^{(4)[1]}_{\kappa}(p,-p,l,-l)}=\nonumber\\&&u^2\int
\frac{d^dq}{(2\pi)^d} \partial_\kappa R_\kappa(q) G^2_0(\kappa;q)
\left\{ G_0(\kappa;q)+  G_0(\kappa;p+l+q)+
G_0(\kappa;p-l+q)\right\},
\end{eqnarray}
which can be integrated easily to give
\begin{eqnarray}\label{4point1}
&&\!\!\!\!\!\!\!\Gamma^{(4)[1]}_{\kappa}(p,-p,l,-l)=\nonumber\\
&&u\!-\! \frac{u^2}{2}\int \frac{d^dq}{(2\pi)^d}  G_0(\kappa;q)
 \left\{ G_0(\kappa;q)+  G_0(\kappa;p+l+q)+ G_0(\kappa;p-l+q)\right\} ,\nonumber \\
\end{eqnarray}
where we imposed the initial condition
$\Gamma^{(4)}_{\kappa=\Lambda}(p,-p,l,-l,\rho=0)=u$ (the integrand
in eq.~(\ref{4point1})  should, for finite $\Lambda$, be subtracted
from its value at $\kappa=\Lambda$ in order to satisfy this initial
condition; the corresponding contribution, however, vanishes in the
limit $\Lambda\to\infty$, and we assume here that $\Lambda$ is large
enough so that it can be neglected.)

Going now to  2-loop, one puts in the r.h.s. of eq.~(\ref{gamma2champnonnul}) the 1-loop expressions of
both the propagator and the 4-point functions. Since we are interested only in the momentum dependence of
the 2-point function, we consider only  the terms in the flow equation that depend on $p$, i.e.,  $\Sigma_\kappa(p)=\Gamma^{(2)}_\kappa(p)-\Gamma^{(2)}_\kappa(0)-p^2$.
Since the momentum dependent  terms
originate entirely from the contribution of order $u^2$ in  $\Gamma^{(4)[1]}$,  we can use $G_0$ as propagator.
We have therefore \begin{eqnarray}\label{2loops}
\partial_\kappa \Sigma_\kappa^{[2]}(p)=
\frac{u^2}{2}\int \frac{d^dl}{(2\pi)^d} \partial_\kappa
R_\kappa(l)G_0^2(\kappa;l)
 \int \frac{d^dq}{(2\pi)^d} G_0(\kappa;q) (G_0(\kappa;p+l+q)\!-\!G_0(\kappa;l+q)).\nonumber\\
\end{eqnarray}
This expression can also be integrated to give
\begin{eqnarray}\label{DeltaGamma2a}
\Sigma_{\kappa}^{[2]}(p) =-\frac{u^2}{6} \int \frac{d^dl}{(2\pi)^d}
\int \frac{d^dq}{(2\pi)^d} G_0(\kappa;l) G_0(\kappa;q)
(G_0(\kappa;p+l+q)-G_0(\kappa;l+q)).
\end{eqnarray}

At this point, we need to deal with the fact that the 2-loop expression for the self-energy is IR divergent. And indeed when $\kappa\to 0$ at fixed $p$, the integral in eq.~(\ref{DeltaGamma2a}) diverges. In order to go around this difficulty, we consider the derivative $\partial_p \Delta \Gamma^{(2)}_\kappa(p)$
\begin{eqnarray}
\frac{\partial\Sigma_{\kappa}^{[2]}(p)}{\partial |p|}=
\frac{u^2}{3}\!\int \frac{d^dl}{(2\pi)^d} \int \frac{d^dq}{(2\pi)^d}
G_0(l)
G_0(q) G_0^2\;( p\!+\!l\!+\!q)({\bf l\!+\!q\!+\!p}).\hat{{\bf p}}\; (1+ R_\kappa'(l\!+\!p\!+\!q)),\nonumber\\
\end{eqnarray}
where $\hat{{\bf p}}\equiv {\bf p}/|{\bf p}|$ and $R_\kappa'(q)\equiv \partial_{q^2}R_\kappa(q)$. The limit $\kappa \to 0$ can now be taken, and yields
\begin{eqnarray}
\frac{\partial\Sigma_{\kappa=0}^{[2]}(p)}{\partial |p|}
=\frac{u^2}{6}\int \frac{d^dq}{(2\pi)^d}\frac{2q.\hat{p}}{q^4} \int
\frac{d^dl}{(2\pi)^d} \frac{1}{l^2}\frac{1}{(l+p-q)^2} .
\end{eqnarray}
Performing the  integral over $l$ and those over $\cos \theta=\hat
{\bf p} . \hat {\bf q}$ and $|{\bf q}|$, one recovers  the well
known result ( in $d=3$):
\begin{eqnarray}\label{finexact}
\frac{\partial\Sigma_{\kappa=0}^{[2]}(p)}{\partial |p|}
&=&\frac{1}{24}\frac{u^2}{(2\pi)2}\int_0^\infty \frac{d|q|}{|q|}
\int_0^\pi d\theta \frac{\sin \theta\, \cos \theta}{(p^2+q^2-2|p|
|q| \cos \theta )^{1/2}}
\nonumber\\
&=&\frac{u^2}{96\pi^2} \frac{1}{|p|}.
\end{eqnarray}

Let us now turn to  the perturbative limit of eq.~(\ref{2pointclosed}). Note that, at both 0- and 1-loop
orders, the predictions of eqs.~(\ref{gamma2champnonnul}) and (\ref{2pointclosed}) for the self-energy
coincide. A difference arises at 2-loop order since, at the LO of the approximation scheme, we should insert
in  eq.~(\ref{2loops})  $\Gamma^{(4)[1]}_{\kappa}(p,-p,0,0)$ instead of $\Gamma^{(4)[1]}_{\kappa}(p,-p,l,-l)$
as we did in the exact calculation, where the expression of $\Gamma^{(4)[1]}$ is given in eq.~(\ref{4point1}).
That is, the LO flow equation reads
\begin{eqnarray}
\partial_\kappa \Sigma_{\kappa}^{[2]LO}(p)=
\frac{u^2}{2} \int \frac{d^dl}{(2\pi)^d} \partial_\kappa
R_\kappa(l)G_0^2(\kappa;l)
 \int \frac{d^dq}{(2\pi)^d} G_0(\kappa;q) (G_0(\kappa;p+q)-G_0(\kappa;q))
 .\nonumber\\
\end{eqnarray}
In contrast to what happens with eq.~(\ref{2loops}), here the
integration over $\kappa$ can no longer be done analytically and we
have to deal with a third integral. Let us call $\kappa'$ the
variable of this integration, and integrate over
$t'=\log(\kappa'/|p|)$. After making the changes of variables $q \to |p|
q$ and $l \to |p|l$, one obtains:
\begin{eqnarray}
&&\Sigma_{\kappa}^{(2)[2]LO}(p)=-\frac{u^2}{2} |p|^{2(d-3)} \nonumber\\
&&\times \int_{\log \kappa/|p|}^\infty
 dt \int \frac{d^dl}{(2\pi)^d} \partial_{t'}R_{\kappa'}(l)
G_0^2(\kappa;l) \int \frac{d^dq}{(2\pi)^d} G_0(\kappa;q) (G_0(\kappa;\hat{p}+q)-G_0(\kappa;q)) .\nonumber\\
\end{eqnarray}
Now, the derivative with respect to $|p|$ is very simple
because, in $d=3$, it only enters in the integration limit. One has:
\begin{eqnarray}
\frac{\partial{\Sigma_{\kappa}^{(2)[2]LO}(p)}} {\partial
|p|}=-\frac{u^2}{2|p|}   \int \frac{d^3l}{(2\pi)^3}
\partial_t R_\kappa(l)G_0^2(\kappa;l) \int \frac{d^3q}{(2\pi)^3}
G_0(\kappa;q)
 (G_0(\kappa;\hat{p}+q)-G_0(q)),\nonumber\\
&&
\end{eqnarray}
where $t=\log(\kappa/|p|)$. In can be verified that the first term, i.e.,
that containing $\hat p$, vanishes when $\kappa \ll |p|$. In
this limit:
\begin{eqnarray}\label{finLO}
\frac{\partial{\Sigma_{\kappa=0}^{[2] LO}(p)}}{\partial
|p|} &=& \frac{1}{2}\frac{u^2}{|p|}\int \frac{d^3l}{(2\pi)^3}
\partial_t R_\kappa(l)G_0^2(\kappa;l)
 \int \frac{d^3q}{(2\pi)^3} G^2_0(\kappa;q)\nonumber\\
&=&\frac{u^2}{9 \pi^4}\frac{1}{|p|}.
\end{eqnarray}

Comparing eqs.~(\ref{finexact}) and (\ref{finLO}) one sees that they
both predict a logarithmic behavior for the self-energy, the ratio
of their respective coefficients being:
\beq
\frac{1/(9\pi^4)}{1/(96\pi^2)} \simeq 1.08.
\eeq

\bibliographystyle{unsrt}

\end{document}